\documentclass[twocolumn]{aastex701}

\usepackage{booktabs}
\usepackage{subcaption}
\usepackage{amsmath} 
\usepackage{array}
\usepackage{listings}
\usepackage{xcolor}
\usepackage{CJK}
\usepackage{enumitem}
\usepackage[T1]{fontenc}
\usepackage{ae,aecompl}
\usepackage[percent]{overpic}
\usepackage{graphicx}
\usepackage{upgreek}
\usepackage{amssymb}
\usepackage{caption}
\usepackage{amsmath}
\newcommand{\HI}{\text{H\,{\scshape i}}}
\usepackage{hyperref}

\begin{document}
\begin{CJK*}{UTF8}{gbsn}

\title{FAST Observations of Filamentary and Compact H\,{\sc i} Structure in the Magellanic Stream IV Field}

\correspondingauthor{Qingzheng Yu (余清正)}
\email{qingzheng.yu@unifi.it}
\correspondingauthor{Taotao Fang (方陶陶)}
\email{fangt@xmu.edu.cn}

\author[0009-0001-2677-5336]{Yuxin Xiao (肖雨欣)}
\affiliation{Department of Astronomy, Xiamen University, Xiamen, Fujian 361005, People's Republic of China}
\email{xiaoyuxin@stu.xmu.edu.cn}

\author[orcid=0000-0003-3230-3981]{Qingzheng Yu (余清正)}
\email{qingzheng.yu@unifi.it}
\affiliation{Dipartimento di Fisica e Astronomia, Universit\`a degli Studi di Firenze, Via Giovanni Sansone 1, 50019 Sesto Fiorentino, Firenze, Italy}
\affiliation{INAF - Osservatorio Astrofisico di Arcetri, Largo Enrico Fermi 5, I-50125 Firenze, Italy}

\author[0000-0002-2853-3808]{Taotao Fang (方陶陶)}
\affiliation{Department of Astronomy, Xiamen University,  Xiamen, Fujian 361005, People's Republic of China}
\email{fangt@xmu.edu.cn}

\author[0000-0003-3010-7661]{Di Li\begin{CJK}{UTF8}{bsmi}
(李菂)
\end{CJK}}
\affiliation{New Cornerstone Science Laboratory, Department of Astronomy, Tsinghua University, Beijing 100084, People’s Republic of China}
\affiliation{State Key Laboratory of Radio Astronomy and Technology, National Astronomical Observatories, Chinese Academy of Sciences, Beijing 100101, People’s Republic of China}
\email{dili@tsinghua.edu.cn}

\author[orcid=0000-0002-3940-2950]{Shulan Yan (鄢淑澜)}
\email{yansl@stu.xmu.edu.cn}
\affiliation{Department of Astronomy, Xiamen University, Xiamen, Fujian 361005, People's Republic of China}

\author[0009-0002-4156-3718]{Shuo Zhang (张硕)}
\affiliation{Department of Astronomy, Shanghai Jiao Tong University, Shanghai 200240, People's Republic of China}
\email{zhangshuo01@sjtu.edu.cn}

\begin{abstract}
The Magellanic Stream (MS) is believed to have formed from gas removed from the Large and Small Magellanic Clouds through tidal forces and hydrodynamic interactions with the Milky Way's gaseous halo. It provides an important laboratory for studying how stripped gas fragments, mixes, and evolves in a circumgalactic environment. In this work, we present H\,{\sc i} observations of a $3.8^\circ\times2.2^\circ$ field in the MS IV region, using the data from the Commensal Radio Astronomy FasT Survey (CRAFTS). The total H\,{\sc i} mass in the analyzed field is $\simeq 5.3 \times 10^{6} (d/120\,{\rm kpc})^{2}\,M_{\odot}$, where $d$ is the distance to the MS IV gas. The data resolve the emission into three coherent filamentary H\,{\sc i} structures with related but distinct velocity trends. To characterize the H\,{\sc i} structures and study potential multiphase gas, we adopt a Gaussian decomposition procedure to identify and reconstruct sources. Our results indicate that the field is dominated by one major filamentary H\,{\sc i} complex, together with several smaller kinematic clump-like components. Among these identified sources, only one source likely shows multiple velocity components. Because several sources appear spatially overlapped in projection, we further examine their apparent overlap regions using position-velocity (P-V) diagrams. P-V diagrams across the apparent overlap region show no clear intermediate-velocity bridge or V-shaped structure, favoring line-of-sight projection over direct cloud-cloud collision. These results demonstrate the value of deep, high-angular-resolution H\,{\sc i} observations for resolving faint emission, compact morphologies, and kinematic structure in the MS.
\end{abstract}


\section{Introduction} 

The Magellanic Stream (MS) is an extended system of neutral and ionized gas surrounding the two most massive satellite galaxies of the Milky Way (MW): the Large and Small Magellanic Clouds (LMC and SMC). It is currently believed to be created by tidal forces, ram pressure, and halo interactions \citep{2012MNRAS.421.2109B, 2019BAAS...51c..20F}. Earlier studies have established that the neutral MS spans approximately $140^\circ$ in length and $10^\circ$ in width, divided into six segments (MS I$-$VI) \citep{1971A&A....12...59D, 1974ApJ...190..291M, 2010ApJ...723.1618N, 2015ApJ...813..110H}, with a total H\,{\sc i} mass of about $2.8 \times 10^{8} M_{\odot}$, including the Interface Region \citep{2005A&A...432...45B, 2010ApJ...723.1618N}. Observations reveal that the central portion of the MS can be divided into two primary filaments \citep{1982MNRAS.199..281C,1983AJ.....88...62M,2008ApJ...679..432N,2013ApJ...772..111R}. The MS is the only known gaseous tidal stream in the vicinity of the MW. It dominates all other gaseous high-velocity clouds (HVCs) in the Galactic halo, both in terms of gas mass and inflow rate. Hydrodynamic instabilities and thermal conduction likely lead to partial evaporation and mixing into the hot halo \citep{2014ApJ...787..147F, 2012ARA&A..50..491P}, while denser clumps may survive and eventually re-accrete onto the Galactic disk, contributing to the gas supply for future star formation \citep{2015ApJ...813...94T, 2017ASSL..430..323F}. Thus, the MS provides an important nearby laboratory for studying the structure, kinematics, and survival of circumgalactic gas in the MW halo.

\par The MS also contains abundant small-scale structures, which predominantly exist as discrete H\,{\sc i} clouds exhibiting compact and multiphase properties \citep{1979AJ.....84.1173H,1979MNRAS.186..433M,2002ApJ...576..773S,2008ApJ...680..276S,2003ApJ...586..170P}. The typical H\,{\sc i} linewidth in the MS is about 25$-$30 km s$^{-1}$ \citep{1984IAUS..108..125M}, while Gaussian decomposition studies have identified much narrower components in selected regions, with linewidths as small as $\sim3$ km s$^{-1}$ \citep{2006A&A...455..481K,2008ApJ...680..276S}. Such linewidths may indicate cooler neutral gas if they are dominated by thermal broadening, but turbulence, unresolved velocity gradients, and line blending can also contribute to the observed linewidths. Therefore, high-sensitivity and high-resolution H\,{\sc i} observations are needed to characterize the local morphology and kinematics of compact Stream structures. In regions where multiple velocity components overlap on the sky, position-velocity (P-V) diagrams have been used for diagnosing kinematics of gas clouds such as line-of-sight projection or possible physical connection, e.g., direct cloud-cloud collision may produce intermediate-velocity bridges or V-shaped structures in P-V space \citep{2021PASJ...73S...1F}. 

\par Previous H\,{\sc i} surveys provide important context for the structure of the MS. The H\,{\sc i} Parkes All Sky Survey (HIPASS; with an angular resolution of 15.5$'$) conducted by \citet{2003ApJ...586..170P} and \citet{2005A&A...432...45B}  revealed that near Declination $\sim0^\circ$, the main filaments of the MS fragment into numerous clumps and filaments, ultimately presenting a chaotic morphology at the tip. Using data from the Leiden-Argentine-Bonn (LAB) all-sky H\,{\sc i} survey (with a spatial resolution of 36$'$), \citet{2008ApJ...679..432N} showed that at least half of the MS originated in the LMC, with an age of $\sim$ 1.74 Gyr. \citet{2002ApJ...576..773S} and \citet{2008ApJ...680..276S} showed that the small-scale structure at the northern end of the MS consists of clumps and occasional filaments with the Arecibo L-band Feed Array (GALFA; with an angular resolution of 3.5$'$). Several clumps exhibit distinct head-tail morphologies, suggesting that the MS gas is interacting with the Galactic halo gas. \citet{2009ApJS..181..398M} used Parkes Galactic All-Sky Survey (GASS) data (effective angular resolution 16$'$) to clearly reveal fine structure in the MS down to scales of 30$'$. \citet{2018MNRAS.474..289W} created the most complete all-sky survey of Galactic H\,{\sc i} emission with an angular resolution of 16.2$'$, based on the new HI4PI all-sky ($4\pi$ sr)  H\,{\sc i} survey \citep{2016A&A...594A.116H}. The Galactic ASKAP Survey (GASKAP) is designed to map H\,{\sc i} and OH emission in the Milky Way and the Magellanic System with high angular and spectral resolution \citep{2013PASA...30....3D}. Future ASKAP/GASKAP comparisons will provide an important interferometric reference for separating compact H\,{\sc i} structures from more extended diffuse emission across 
angular scales. 

\par In this work, we use the available data from the Commensal Radio Astronomy FasT Survey (CRAFTS) to investigate how the MS is structured on small (few-arcminute) scales in a specific region. CRAFTS is a legacy survey conducted with the Five-hundred-meter Aperture Spherical radio Telescope (FAST) \citep{2018IMMag..19..112L,2019SCPMA..6259506Z}. The original beam size of CRAFTS is about 2$\farcm$9 while the effective beam size after reduction is 4$'$. The higher angular resolution of CRAFTS allows us to resolve finer H\,{\sc i} structures in this field than those seen in HI4PI. The currently released CRAFTS data cover only a limited and non-contiguous part of MS IV. We therefore treat the present field as a local study of the main emission complex rather than as a representative map of the entire Stream. The present field also serves as a pilot study for extending this analysis to the entire MS tail after the completion of the CRAFTS survey. We use CRAFTS to examine the small-scale H\,{\sc i} morphology and kinematics in this MS IV field, including whether the emission is organized into filamentary structures and compact clumps. We also test whether the apparent spatial overlap between velocity components is accompanied by P-V signatures expected for direct cloud-cloud collision, or is more consistent with line-of-sight projection.

\par In Section~\ref{sec:data}, we briefly outline our observing and data processing methods. In Section~\ref{sec:results}, we present fundamental results derived from the data analysis. In Section~\ref{sec:discussion}, we discuss the structure of the MS and its implications for the Milky Way, based on our observational findings. In Section~\ref{sec:Systematics}, we discuss the main systematic uncertainties and robustness tests. Finally, in Section~\ref{sec:Conclusions}, we summarize the main conclusions of this study.

\begin{figure*}[ht!]
\centering
\includegraphics[width=\textwidth]{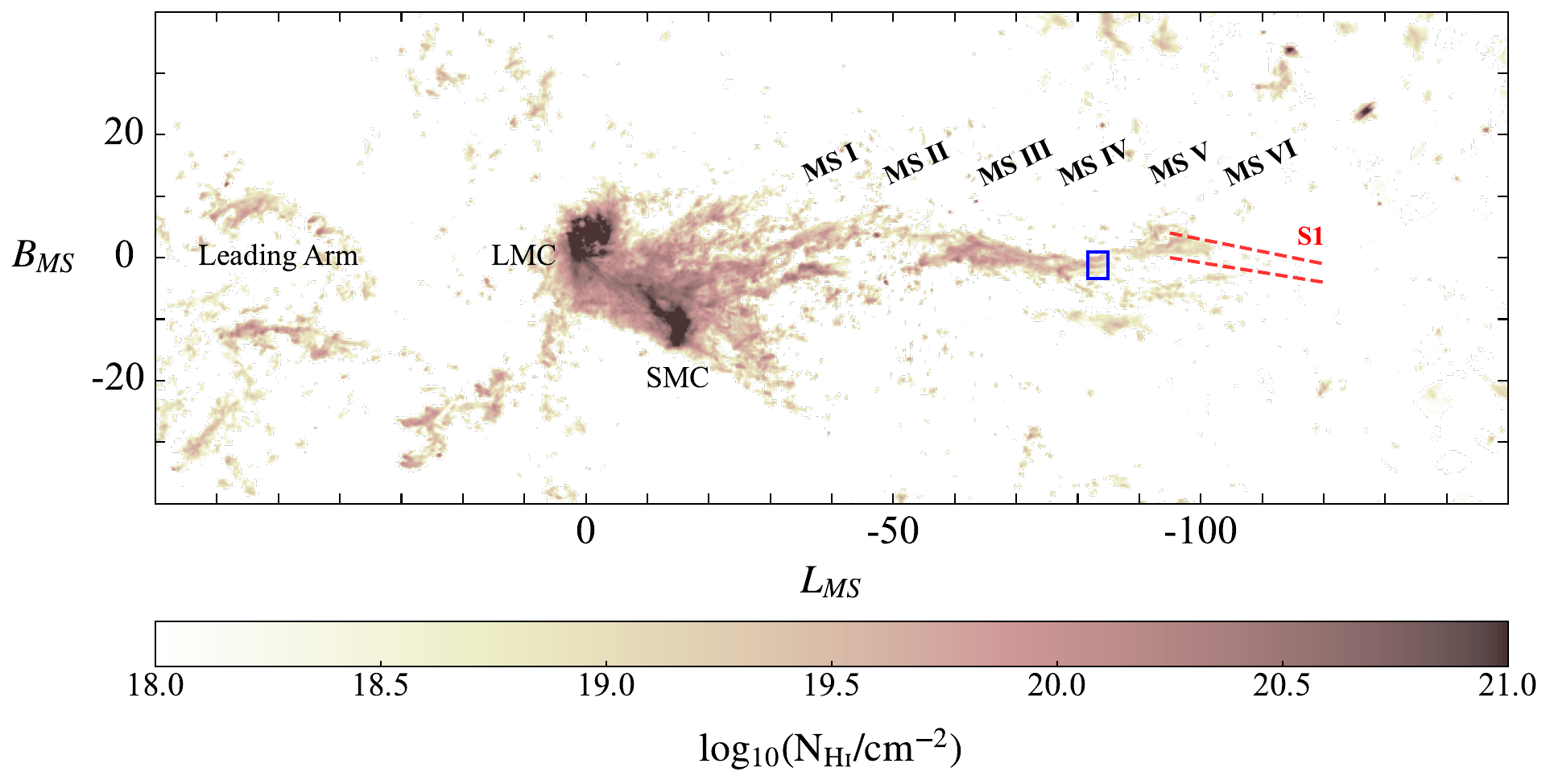}
\caption{H\,{\sc i} column density distribution of MS, in the range of $\mathrm{log}_{10}(\mathrm{N}_{\mathrm{\HI}}/\mathrm{cm}^{-2})$ = 18 to 21 with the beam size of 16.2$'$ from the HI4PI survey. The emission is shown in the Magellanic coordinate system. The blue box marks the region analyzed in this work. The red dotted line marks the range of ``S1'' of \citet{2008ApJ...680..276S}.
\label{fig:all}}
\end{figure*}

\section{Data and Methods} \label{sec:data}

\subsection{Observations} \label{subsec:cubes}
We used the publicly available data from the CRAFTS survey project of FAST. FAST adopts a drift scan mode, which is a survey observation strategy with a fixed pointing that utilizes Earth's rotation to scan the sky. This mode employs a 19-beam L-band receiver rotated by $23.4^\circ$, covering declinations from $-14^\circ$ to $+66^\circ$. The drift scan mode offers extremely high sensitivity and the capability to accommodate multiple scientific objectives simultaneously. CRAFTS is one of the priority major projects of FAST, which aims to obtain data simultaneously for pulsar search, detection of H\,{\sc i} galaxies, H\,{\sc i} imaging, and radio transients through multiple back ends. CRAFTS has thus far observed approximately 20\% of the planned sky coverage. 

\par The complete map of the MS \citep{2018MNRAS.474..289W} plotted in the Magellanic coordinate system of \citet{2010ApJ...723.1618N} can be seen in Figure~\ref{fig:all}, where the area with a blue border is the scope of our research data in this paper. The public CRAFTS data considered here nominally cover
$350.0^\circ \lesssim \mathrm{RA} \lesssim 360.0^\circ$ and
$-13.0^\circ \lesssim \mathrm{Dec} \lesssim -3.0^\circ$.
However, significant MS H\,{\sc i} emission in the relevant
velocity range is concentrated within
$-6.9^\circ \lesssim \mathrm{Dec} \lesssim -4.7^\circ$, while the
CRAFTS coverage is incomplete at
$357.0^\circ \lesssim \mathrm{RA} \lesssim 360.0^\circ$.
We therefore restrict our analysis to the contiguous main emission
complex spanning
$353.0^\circ \lesssim \mathrm{RA} \lesssim 356.8^\circ$, which
defines the final $3.8^\circ \times 2.2^\circ$ field.
In the Magellanic coordinate system, this corresponds to
$-84.9^\circ \lesssim L_{\rm MS} \lesssim -81.6^\circ$ and
$-3.4^\circ \lesssim B_{\rm MS} \lesssim 1^\circ$ \citep{2010ApJ...723.1618N}.
\par The CRAFTS data used in this work are publicly available from the Science Data Bank \citep{Li2024CRAFTSdata}\footnote{\url{https://www.scidb.cn/detail?dataSetId=7fe34782f6ee4284aa22fc68ab8ab6cd}}. We also use the corresponding HI4PI data for comparison. The HI4PI data are publicly available from the VizieR repository \citep{vizier:J/A+A/594/A116}\footnote{\url{https://cdsarc.cds.unistra.fr/ftp/J/A+A/594/A116/CUBES/EQ2000/CAR/}}. The data-processing and analysis code used in this work is publicly
available on GitHub\footnote{\url{https://github.com/astroyxyxyx/HVC}}
and is permanently archived on Zenodo as version v1.0.0
\citep{xiao_2026_22040479}.

\subsection{Reduction and calibration} \label{subsec:reduction}
The main steps in the CRAFTS data reduction include calibration, radio-frequency interference (RFI) flagging, baseline subtraction, coordinate projection, and spatial/spectral regridding, ultimately producing an RA-Dec-velocity H\,{\sc i} cube suitable for scientific analysis \citep{2018IMMag..19..112L}. We inspected the baseline of the CRAFTS data and corrected it using a first-order polynomial fit. We have tested the uncertainty induced baseline correction and found it to be $2\%$$-$$5\%$ of the integrated flux. Considering uncertainties from flux calibration and baseline subtraction in CRAFTS' pipeline, we adopt a 10\% systematic uncertainty for the following analysis based on previous studies \citep{2025ApJS..279...32Y,2025RAA....25a5011X}. The final data cube corrected with the first-order polynomial fit was adopted for the analysis presented in this paper. The coordinate resolution is 0.025$^\circ$, velocity resolution is 0.20 km s$^{-1}$ \citep{2018IMMag..19..112L}.

We estimated the rms noise using the same spatial region as the science field, but over an emission-free velocity interval of
$-400~{\rm km~s^{-1}} \lesssim V_{\rm LSR} \lesssim -350~{\rm km~s^{-1}}$.
The CRAFTS cube has an rms noise of $\sigma_{\rm rms}\approx120$ mK per 0.2 km s$^{-1}$ channel, with an effective beam size of $4'$.
The $1\sigma$ H\,{\sc i} column-density sensitivity over a linewidth $\Delta v$ is estimated as
\begin{equation}
\sigma_{N_{\rm HI}}
=
1.8\times10^{18}
\,\sigma_{\rm rms}\,\delta v
\sqrt{\frac{\Delta v}{\delta v}}
\ {\rm cm^{-2}},
\label{eq:nhi_sens}
\end{equation}
where $\sigma_{\rm rms}$ is the rms noise per channel in K, $\delta v$ is the channel width, and $\Delta v$ is the assumed linewidth.
For $\delta v=0.2~{\rm km~s^{-1}}$ and $\Delta v=20~{\rm km~s^{-1}}$, this corresponds to a $5\sigma$ sensitivity of
$N_{\rm HI}\approx2.2\times10^{18}~{\rm cm^{-2}}$ for beam-filling emission
\citep{1990ARA&A..28..215D,2024SCPMA..6719511Z}.

To compare the CRAFTS and HI4PI data directly, we smoothed the CRAFTS data to the angular and spectral resolution of HI4PI. The HI4PI data have an rms noise of $\sigma_{\rm rms}\approx40$ mK per 1.3 km s$^{-1}$ channel at a beam size of $16.2'$, corresponding to a $5\sigma$ H\,{\sc i} column-density sensitivity of $N_{\rm HI}\approx1.9\times10^{18}$ cm$^{-2}$ over a 20 km s$^{-1}$ linewidth, for beam-filling emission. After smoothing, the CRAFTS data reach $\sigma_{\rm rms}\approx10$ mK per 1.3 km s$^{-1}$ channel, corresponding to $N_{\rm HI}\approx4.6\times10^{17}$ cm$^{-2}$ at the same significance level and linewidth, for beam-filling emission.

\subsection{Moment maps and masks} \label{subsec:maps}
The original CRAFTS cube spans $-600~{\rm km\,s^{-1}} \leq V_{\rm LSR} \leq
600~{\rm km\,s^{-1}}$. Based on previous observations of this region \citep{2018MNRAS.474..289W,2014ApJ...792...43F}, we searched for MS H\,{\sc i} emission over $-350~{\rm km\,s^{-1}} \lesssim V_{\rm LSR} \lesssim -150~{\rm km\,s^{-1}}$ and found that the emission is concentrated within $-300~{\rm km\,s^{-1}} \lesssim V_{\rm LSR} \lesssim -200~{\rm km\,s^{-1}}$.
\par We processed the noise mask using SoFiA-2, the Source Finding Application described by \citet{2021MNRAS.506.3962W}. SoFiA-2 was originally designed to detect and characterize galaxies in three-dimensional extragalactic H\,{\sc i} data cubes. It also provides automatic flagging, noise normalization, the \textit{smooth and clip} (S+C) source finder, and reliability estimation \citep{2012PASA...29..296S,2012MNRAS.422.1835S,2015MNRAS.448.1922S,2021MNRAS.506.3962W}. For the input data cube, a detection threshold needs to be specified. The S+C finder then smooths the data cube over multiple spatial and spectral scales, and the linker combines pixels that satisfy the specified conditions.

\section{Results} \label{sec:results}
\subsection{Moment maps} \label{subsec:maps}
We visually inspected the detected sources and data cubes to verify the detections. The adopted SoFiA-2 parameters represent a compromise between retaining faint structures and removing noise: scfind.kernels = 0, 1, 2, 3, 4, 5 (spatial Gaussian kernel sizes), scfind.threshold = 4.5 (the flux threshold for source detection), linker.radius = 1 (pixels with a separation of up to this value will be merged into the same source) and linker.minSize = 10 (minimum size of sources in the spatial dimension in pixels). 

\par
\begin{figure*}[t!]
\centering

\begin{minipage}{0.95\textwidth}
\includegraphics[width=\linewidth]{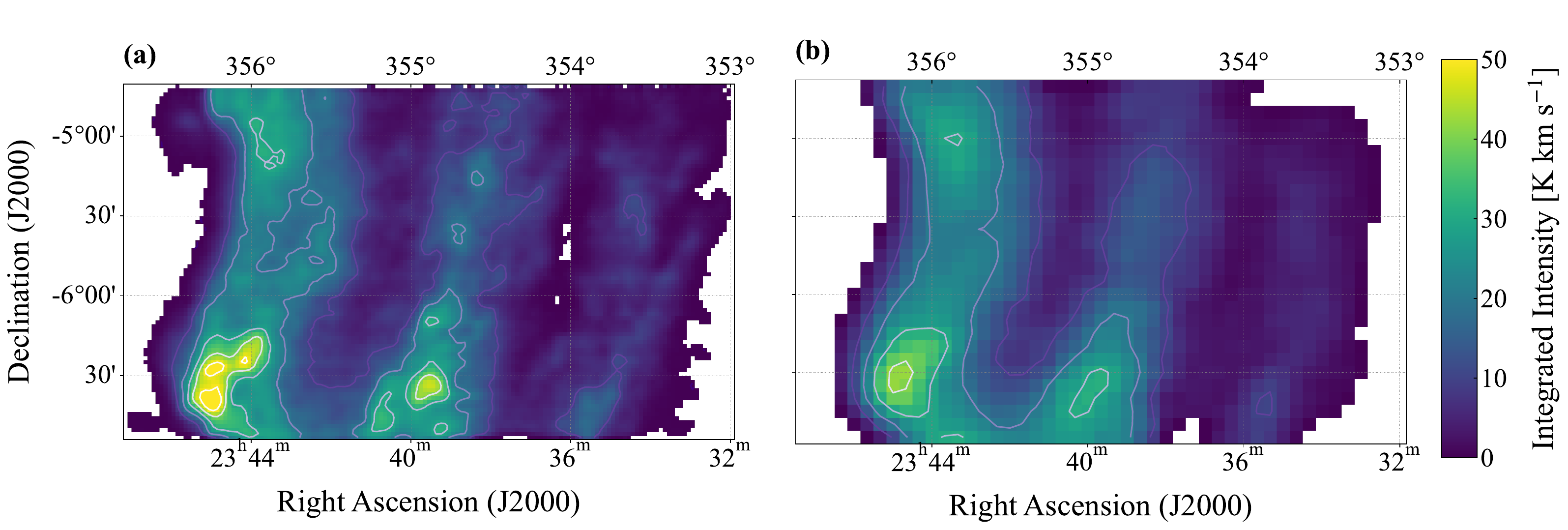}
\end{minipage}

\vspace{0.2em}

\begin{minipage}{0.95\textwidth}
\includegraphics[width=\linewidth]{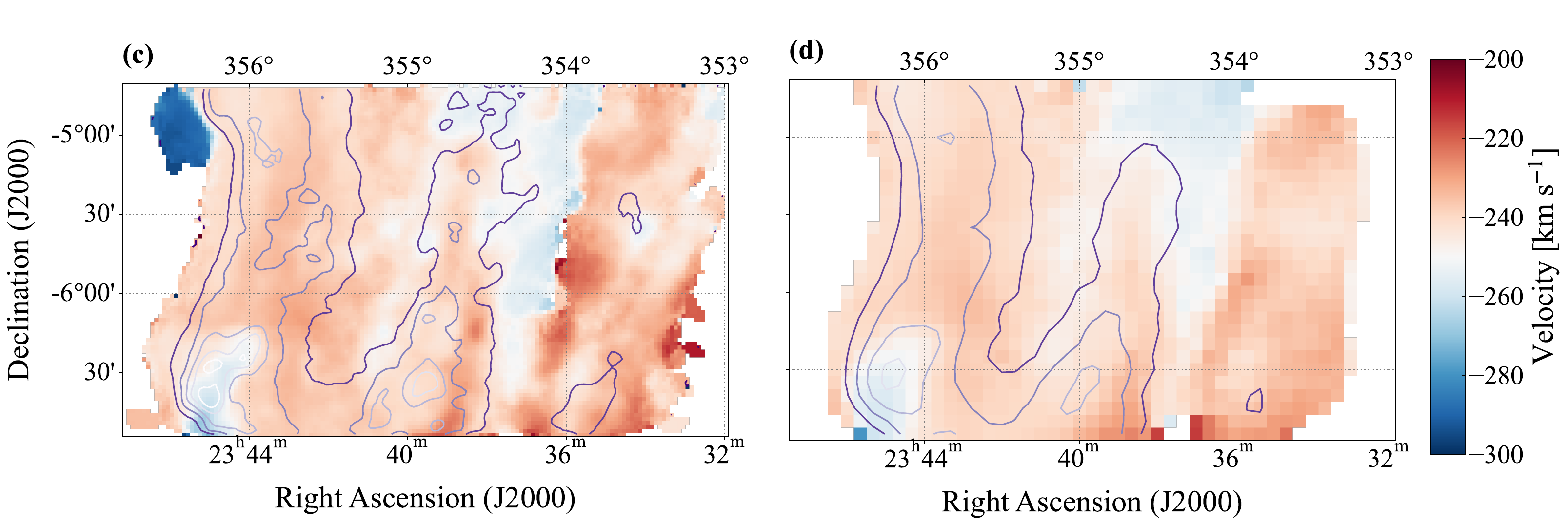}
\end{minipage}

\vspace{0.2em}

\begin{minipage}{0.95\textwidth}
\includegraphics[width=\linewidth]{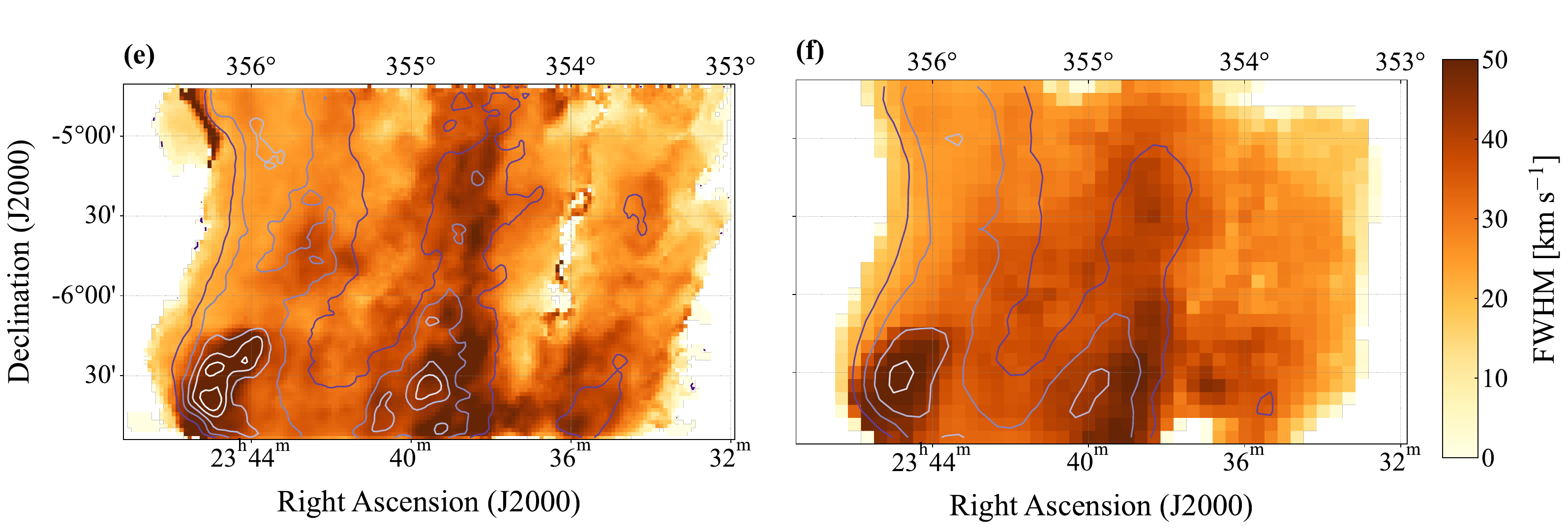}
\end{minipage}

\caption{Comparison of the CRAFTS and HI4PI observations. Panels (a) and (b) show the integrated intensity maps (Moment 0 maps) derived from CRAFTS and HI4PI, respectively. The color scale and contours range from 0 to 50 K km s$^{-1}$, with a contour interval of 10 K km s$^{-1}$. Panels (c) and (d) show the corresponding velocity field maps (Moment 1 maps), with the color scale ranging from $-300$ to $-200$ km s$^{-1}$. Panels (e) and (f) show the corresponding FWHM maps, with the color scale ranging from 0 to 50 km s$^{-1}$. The contours in panels (c)--(f) are the same as those in panels (a) and (b).}
\label{fig:CRAFTS_HI4PI_moments}
\end{figure*}

 The Moment 0, Moment 1, and FWHM maps for the CRAFTS and HI4PI data in the region of $353.0^\circ$ $\lesssim$ RA $\lesssim$ 356.8$^\circ$, $-6.9^\circ$ $\lesssim$ DEC $\lesssim$ $-4.7^\circ$ are shown in Figure~\ref{fig:CRAFTS_HI4PI_moments}. The contour levels representing the integrated intensity are set from 0 to 50 K km s$^{-1}$ in intervals of 10 K km s$^{-1}$. The FWHM map is obtained from the Moment 2 map using 
$\mathrm{FWHM}=2\sqrt{2\ln2}\,\sigma_v \simeq 2.355\sigma_v$.

\par The uncertainties from the output of SoFiA-2 are purely statistical errors, and we considered the systematic errors from flux calibration and baseline correction in the following analysis. After applying the SoFiA-2 mask, the integrated fluxes are $1563.1 \pm 156.3$ Jy km s$^{-1}$ for CRAFTS and $1588.4 \pm 158.8$ Jy km s$^{-1}$ for HI4PI. After smoothing the CRAFTS cube to the angular and spectral resolution of HI4PI and applying the HI4PI mask, the integrated CRAFTS flux becomes $1460.3 \pm 146.0$ Jy km s$^{-1}$, within the HI4PI error range. The error is estimated as:
\begin{equation}
\begin{aligned}
\mathrm{err}_{F_{\mathrm{int}}}
&=
\sqrt{
N_{\mathrm{ch}} \times (\delta v\times \mathrm{\sigma_{\rm spec})}^{2}
+
\left(0.1 F_{\mathrm{int}}\right)^{2}
}, 
\end{aligned}
\end{equation}
$N_\mathrm{ch}$ is the number of velocity channels included in the integrated flux \citep{2021MNRAS.503.5385Z}, $\sigma_{\rm spec}$ is the rms uncertainty per channel of the spatially integrated spectrum after converting from K to Jy and $\delta v$ is the channel width ($\sim 0.2$ km s$^{-1}$).
\par The H\,{\sc i} mass is calculated from the integrated H\,{\sc i} flux using
\begin{equation}
\frac{M_{\rm HI}}{M_{\odot}} =
\frac{2.356 \times 10^{5}}{1+z}
\left(\frac{d}{\rm Mpc}\right)^{2}
\int S(V)\,dV ,
\end{equation}
where $d$ is the unknown distance to the MS IV gas in Mpc, $S(V)$ is the integrated spectral-line profile in Jy, and the velocity integral is in km s$^{-1}$ \citep{2015A&ARv..24....1G}. In this work, the velocity-integrated flux is measured from the CRAFTS data after applying the SoFiA-2 mask, giving
\begin{equation}
F_{\rm int} = \int S(V)\,dV = 1563.1\pm 156.3\ {\rm Jy\,km\,s^{-1}} .
\end{equation}
The distance to the H\,{\sc i} gas in our field has not been directly measured. The classical Stream mass is commonly normalized to 55 kpc, while \citet{2010ApJ...723.1618N} adopted 120 kpc for the extreme Stream tip. \citet{2025ApJ...984..104M} obtained a lower limit of approximately 42 kpc near our field, but this does not uniquely determine the gas distance. We therefore use 120 kpc only as a fiducial normalization and retain the explicit $(d/120\,{\rm kpc})^2$ scaling for all H\,{\sc i} masses. For the region 
$353.0^\circ \lesssim {\rm RA} \lesssim 356.8^\circ$ and 
$-6.9^\circ \lesssim {\rm Dec} \lesssim -4.7^\circ$, the total H\,{\sc i} mass is
\begin{equation}
M_{\rm HI}
\simeq
(5.3 \pm 0.5)\times10^{6}
\left(\frac{d}{120\,{\rm kpc}}\right)^2
M_\odot .
\end{equation}
The error is calculated as:
\begin{equation}
\begin{aligned}
\mathrm{err}_{{M}_\mathrm{HI}}
&=
{M}_\mathrm{HI}
\frac{\mathrm{err}_{F_{\mathrm{int}}}}{F_{\mathrm{int}}}.
\end{aligned}
\end{equation}
For the same sky region and distance scaling, the CRAFTS and HI4PI data give $M_{\rm HI}=(5.3\pm0.5)\times10^6(d/120\,{\rm kpc})^2\,M_\odot$ and $(5.4\pm0.5)\times10^6(d/120\,{\rm kpc})^2\,M_\odot$, respectively. The masses were derived using the respective survey masks but the same mass and uncertainty calculations. Despite differences in angular resolution and masking, the two measurements are consistent within their uncertainties.

\par From the integrated intensity map, we can observe that under the same color scale, the CRAFTS data reveal more detailed structures compared to the HI4PI data, displaying finer features that are not visible in HI4PI. For instance, in the CRAFTS data at RA ($356.5^\circ-355.5^\circ$) and DEC $\sim-6.5^\circ$, the apparent overlap region shows significantly higher integrated intensity and more distinct structures than those in HI4PI. Rather than appearing as a single blurred bright patch, this region in CRAFTS resolves into several bright clumps with clear structural features. At the higher angular resolution of CRAFTS, the emission that appears smoother in HI4PI is resolved into several filamentary and clumpy structures. These results highlight the importance of deep observations with higher angular resolution to reveal fine structures.

\begin{figure*}[ht!]
\centering
\includegraphics[width=\textwidth]{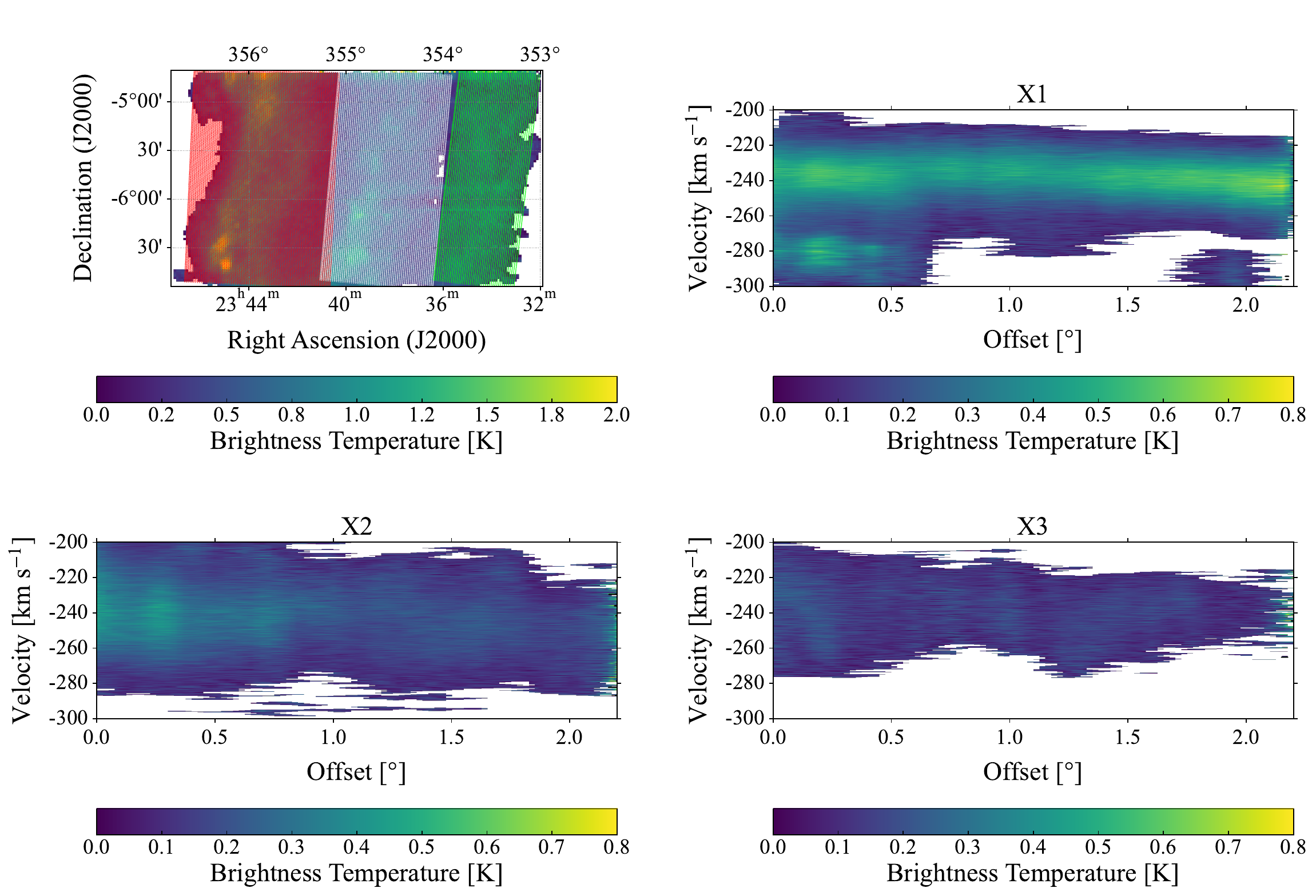}
\caption{Schematic diagram of the three filament directions and the corresponding P-V diagrams. The extraction paths follow the filament directions indicated in the upper-left panel, with widths chosen to encompass the full spatial extent of each filament. The red, white, and green paths correspond to X1, X2, and X3, with integration widths of 90, 70, and 50 arcmin, respectively.
\label{fig:P-V}}
\end{figure*}

\subsection{Resolved morphology} \label{subsec:Resolved morphology}
There are three prominent and coherent large-scale structures noticeable across the whole velocity range. To facilitate further discussion, we define and label the filaments as follows:
\par 1. ``X1'' is centered at RA $\sim356^\circ$, starts at DEC$ \sim -6.9^\circ$ and extends to DEC$ \sim -4.7^\circ$.
\par 2. ``X2'' is centered at RA $\sim354.5^\circ$, starts at DEC$ \sim -6.9^\circ$ and extends to DEC$ \sim -4.7^\circ$, stretching from lower to higher velocities.
\par 3. ``X3'' is centered at RA $\sim353.5^\circ$, starts at DEC$ \sim -6.9^\circ$ and extends to DEC$ \sim -5.0^\circ$. It exhibits a relatively irregular morphology and is comparatively fainter.

\par It can be seen from Figure~\ref{fig:all} that the area studied in this article belongs to MS IV and is connected to more filament-like structures such as MS V. The Moment 0, Moment 1, and FWHM maps show that this part has a clear multi-filamentary structure at the resolution of CRAFTS, which reveals three parallel filamentous structures with different column densities, velocity distributions and velocity dispersions. As shown in Figure~\ref{fig:P-V}, we present a schematic diagram of the three filaments and P-V diagrams along filaments X1, X2, and X3, respectively.

\begin{figure}[ht!]
\centering
\includegraphics[width=\columnwidth]{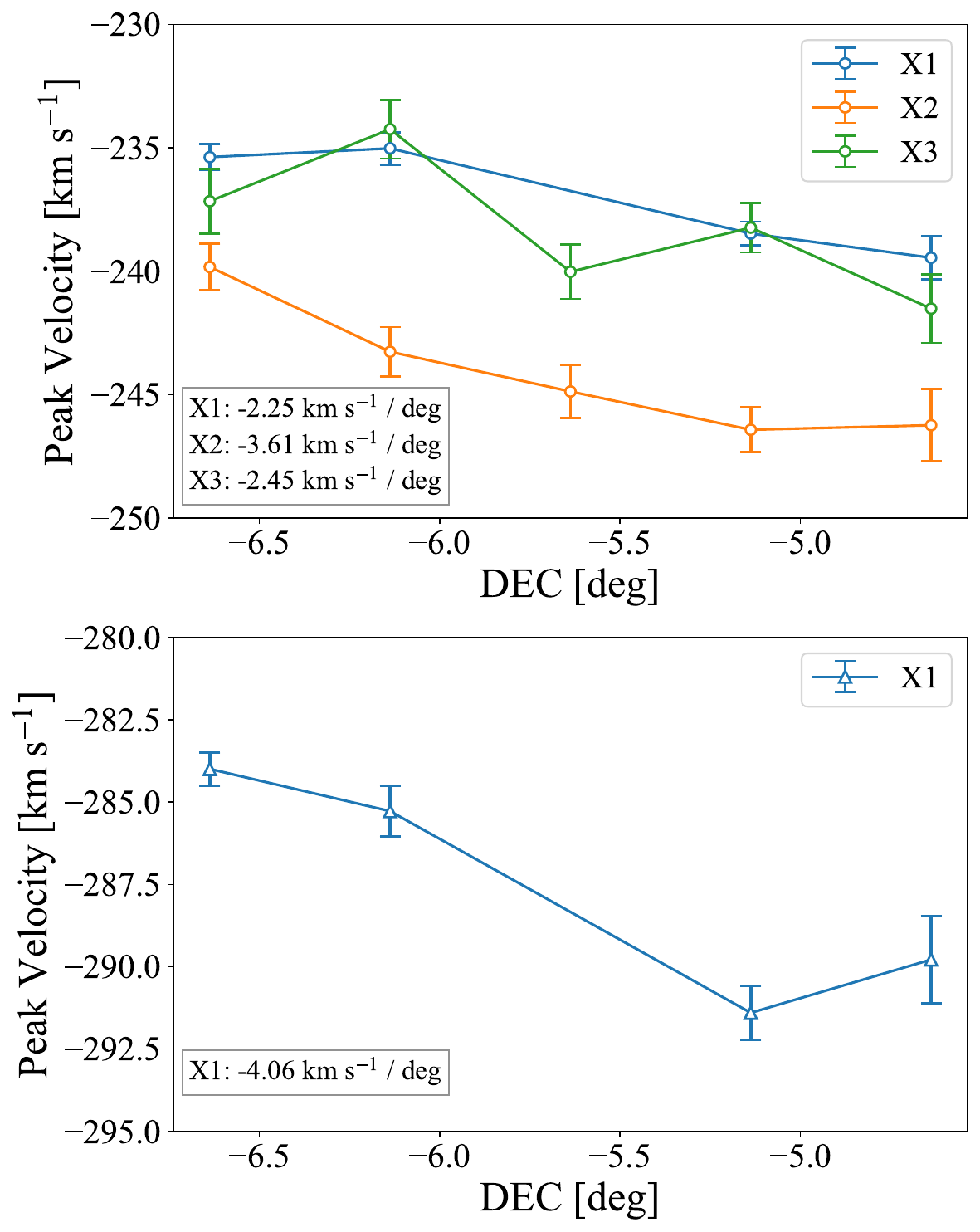}
\caption{Velocity variation as a function of declination. The upper panel shows the mean peak velocities around $-240~\mathrm{km~s^{-1}}$, marked with circles, while the lower panel shows the mean peak velocities around $-280~\mathrm{km~s^{-1}}$, marked with triangles. The velocity uncertainties are estimated from the standard deviation of the peak velocities divided by the square root of the number of independent beams. The same color denotes the same filament.
\label{fig:PeakVelocity}}
\end{figure}

\par We calculated mean velocities in $0.5^\circ$ declination bins along the three filament directions. Within each declination bin, we searched for the peak velocity of the spectrum at each valid pixel around $-240~\mathrm{km~s^{-1}}$ and $-280~\mathrm{km~s^{-1}}$, respectively. The mean of these peak velocities was then adopted as the representative velocity, and the velocity uncertainty was estimated as the standard deviation of the peak velocities divided by the square root of the number of independent beams. The resulting profiles are shown in Figure~\ref{fig:PeakVelocity}. As shown in Figures~\ref{fig:P-V} and~\ref{fig:PeakVelocity}, along the X1 extraction direction, H\,{\sc i} emission is detected near both $\sim -240~\mathrm{km~s^{-1}}$ and $-280$ to $-295~\mathrm{km~s^{-1}}$. We will examine whether this apparent overlap arises from physical interaction or from line-of-sight projection in Section~\ref{subsec:P-V}. Around $-240~\mathrm{km~s^{-1}}$, the three filaments show absolute velocity gradients of approximately $2-4~{\rm km~s^{-1}~deg^{-1}}$, with broadly similar velocity ranges.

\subsection{Gaussian decomposition} \label{subsec:ROHSA}
To study the multiple velocity components of this region, we performed a Gaussian decomposition of the H\,{\sc i} data cube using Regularized Optimization for Hyper-Spectral Analysis (ROHSA; \citealt{2019A&A...626A.101M}). In this approach, each spectrum is modeled as a linear combination of Gaussian components, characterized by their amplitude, centroid velocity, and velocity dispersion. Unlike traditional line-by-line fitting methods, ROHSA exploits the spatial coherence of H\,{\sc i} emission by enforcing smooth variations of the Gaussian parameters across neighboring lines of sight. This is achieved through a regularized non-linear least-squares optimization, allowing for a consistent decomposition even in regions where multiple velocity components overlap along the line of sight  \citep{2019A&A...626A.101M}. 

In the Gaussian decomposition with ROHSA, it is necessary to specify the number of Gaussian components used for the decomposition, N, as well as the strengths of the regularization terms $\lambda_{a}$, $\lambda_{\mu}$, $\lambda_{\sigma}$, and $\lambda'_{\sigma}$ \citep{2019A&A...626A.101M}. In this work, we adopted $\lambda_{a} = 1$, $\lambda_{\mu} = 1$, $\lambda_{\sigma} = 1$, and $\lambda'_{\sigma} = 0$. 

For our dataset, we found that providing reasonable initial guesses for the Gaussian parameters can significantly improve the convergence and stability of the fits: $A$ (amplitude),  $\mu$ (centroid velocity), and $\sigma$ (velocity dispersion). We first performed a visual inspection of the spectra to estimate the number of significant peaks, finding that at least three Gaussian components are required to adequately reproduce the line profiles. We explored models with $N = 3$ to $7$ components, and the corresponding average reduced chi-square values for each $N$ are listed in Table~\ref{tab:chi2}.

\begin{table}[htbp]
\centering
\caption{Average reduced chi-square and its standard deviation for different numbers of Gaussian components.}

\setlength{\tabcolsep}{3pt} 

\begin{tabular}{lccccc}
\toprule
 & $N=3$ & $N=4$ & $N=5$ & $N=6$ & $N=7$  \\
\midrule
Mean reduced $\chi^2$      & 1.040 & 1.039 & 1.039 & 1.041 & 1.041  \\
Standard deviation         & 0.102 & 0.101 & 0.099 & 0.100 & 0.099  \\
\bottomrule
\end{tabular}
\label{tab:chi2}
\end{table}
\par As shown in Table~\ref{tab:chi2}, increasing the number of Gaussian components does not lead to a significant improvement in the average reduced $\chi^2$. We therefore adopt $N = 3$ as the preferred model for the final spectral decomposition, with the centroid velocities initialized at $\mu = -240$ km s$^{-1}$, $-280$ km s$^{-1}$, and $-290$ km s$^{-1}$. 

\par When constructing the catalog, we tested Threshold (minimum signal-to-noise ratio for pixels in the fitted Gaussian component maps) ranging from 1.5 to 3.0. Sources in each Gaussian component were then identified by requiring a minimum size of 10 pixels and allowing a maximum separation of up to 2 pixels between connected regions. We adopted FWHM Range factors of 0.7 and 1.0 \citep{2008ApJ...680..276S}. Two fitted Gaussian components are assigned to the same cataloged source if their centroid-velocity separation is smaller than the adopted FWHM Range factor multiplied by the FWHM of the broader component. Before merging, we visually inspected each source to check whether it corresponded to discontinuous spurious emission. Such false detections were excluded during the merging process. The final source-identification results are summarized in Table~\ref{tab:source_merging_parameters}.

\par In Table~\ref{tab:source_merging_parameters}, $N_{\mathrm{original}}$ represents the total number of sources identified by ROHSA in all Gaussian components before merging. $N_{\mathrm{false}}$ denotes the number of false detections identified by visual inspection, and $N_{\mathrm{merged}}$ is the number of sources remaining after removing the false detections and merging sources according to the adopted FWHM Range criterion. As the threshold increases, the identified sources become more reliable. For Threshold = 2, the number of false detections is relatively small, while relatively faint emission can still be retained. Therefore, we adopt Threshold = 2 and FWHM Range = 0.7 as the fiducial source-identification criterion for the main analysis. 

\begin{deluxetable*}{ccccc}
\tablecaption{Number of identified sources before and after merging under different Threshold and FWHM range settings.
\label{tab:source_merging_parameters}}
\tablehead{
\colhead{Threshold} &
\colhead{FWHM Range} &
\colhead{$N_{\rm original}$} &
\colhead{$N_{\rm false}$} &
\colhead{$N_{\rm merged}$}
}
\startdata
1.5 & 0.7 & 16 & 6 &  10\\
1.5 & 1.0 & 16 & 6 &  7\\
2.0 & 0.7 & 11 & 1 &  9\\
2.0 & 1.0 & 11 & 1 &  7\\
2.5 & 0.7 & 6  & 0 &  6 \\
2.5 & 1.0 & 6  & 0 &  5 \\
3.0 & 0.7 & 7  & 0 &  7 \\
3.0 & 1.0 & 7  & 0 &  6\\
\enddata
\end{deluxetable*}

\par
\begin{figure*}[ht!]
\centering
\includegraphics[width=0.95\textwidth]{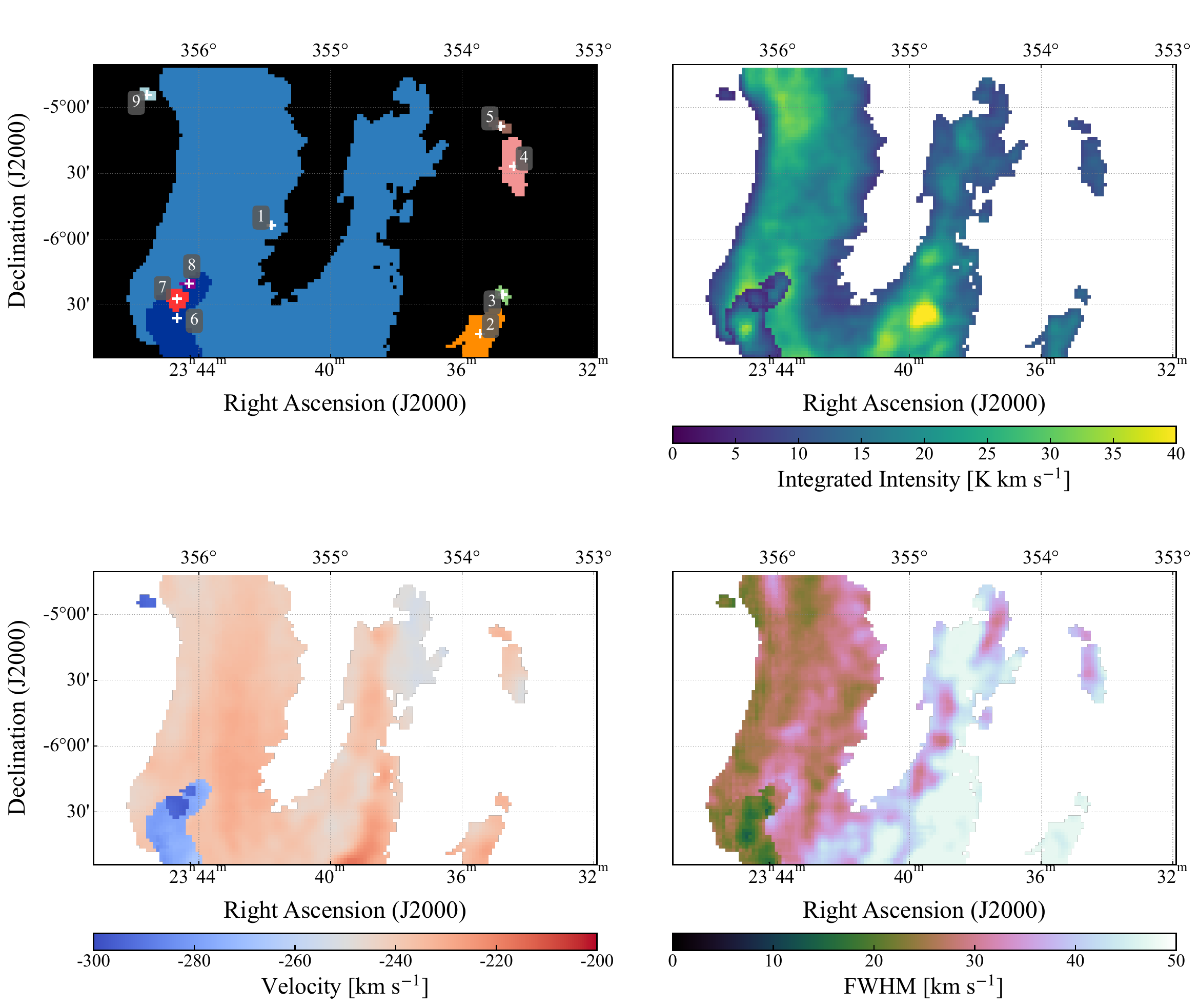}
\caption{Moment 0, Moment 1, and FWHM maps of the 9 cataloged H\,{\sc i} sources for Threshold = 2, FWHM Range = 0.7. All panels are constructed from the ROHSA Gaussian fitting results. The left image in the first row shows the shape and location of the 9 sources, marked with different colors. The Moment 0, Moment 1, and FWHM maps are derived from the corresponding fitted Gaussian components. 
\label{fig:CATALOG}}
\end{figure*}

\begin{figure*}
\centering

\begin{subfigure}{0.45\textwidth}
\centering
\includegraphics[width=\linewidth]{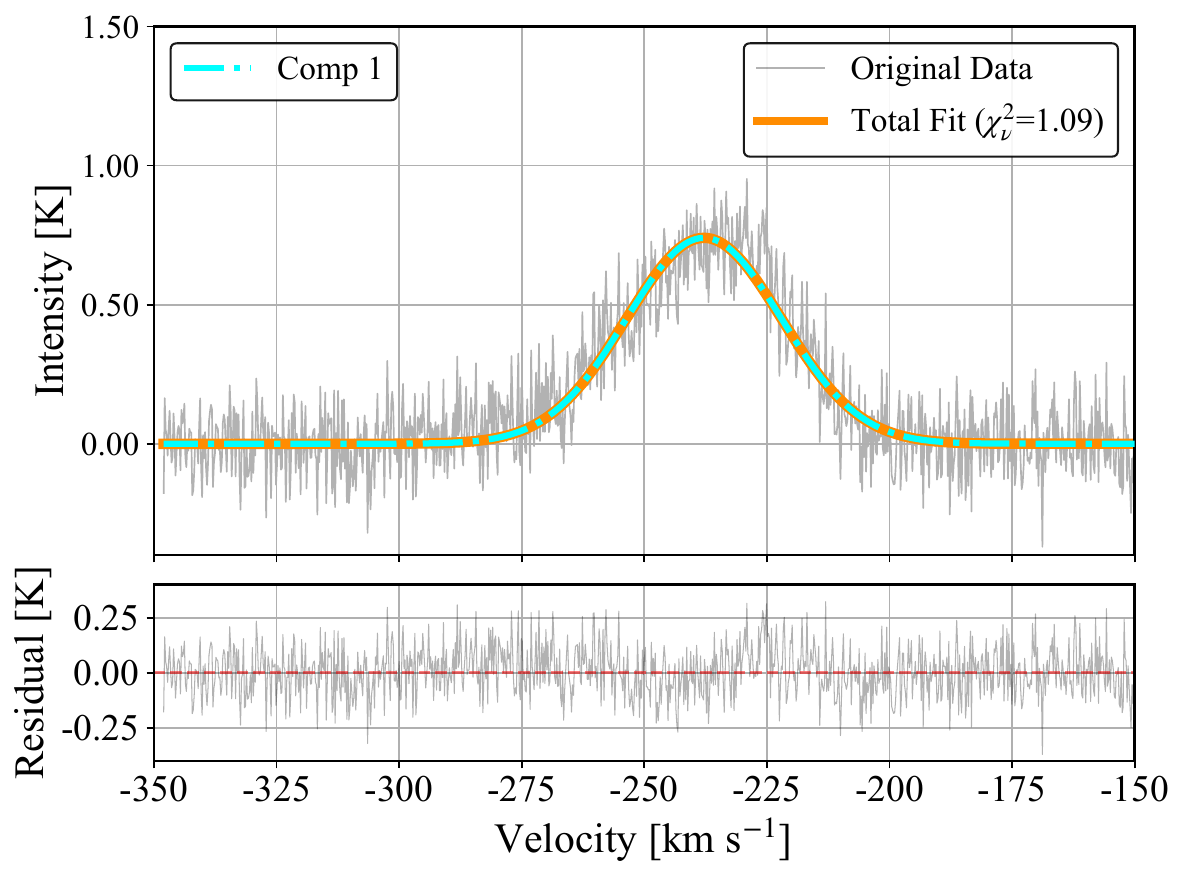}
\label{subfig:spectrum1}
\end{subfigure}
\hspace{0.02\textwidth} 
\begin{subfigure}{0.45\textwidth}
\centering
\includegraphics[width=\linewidth]{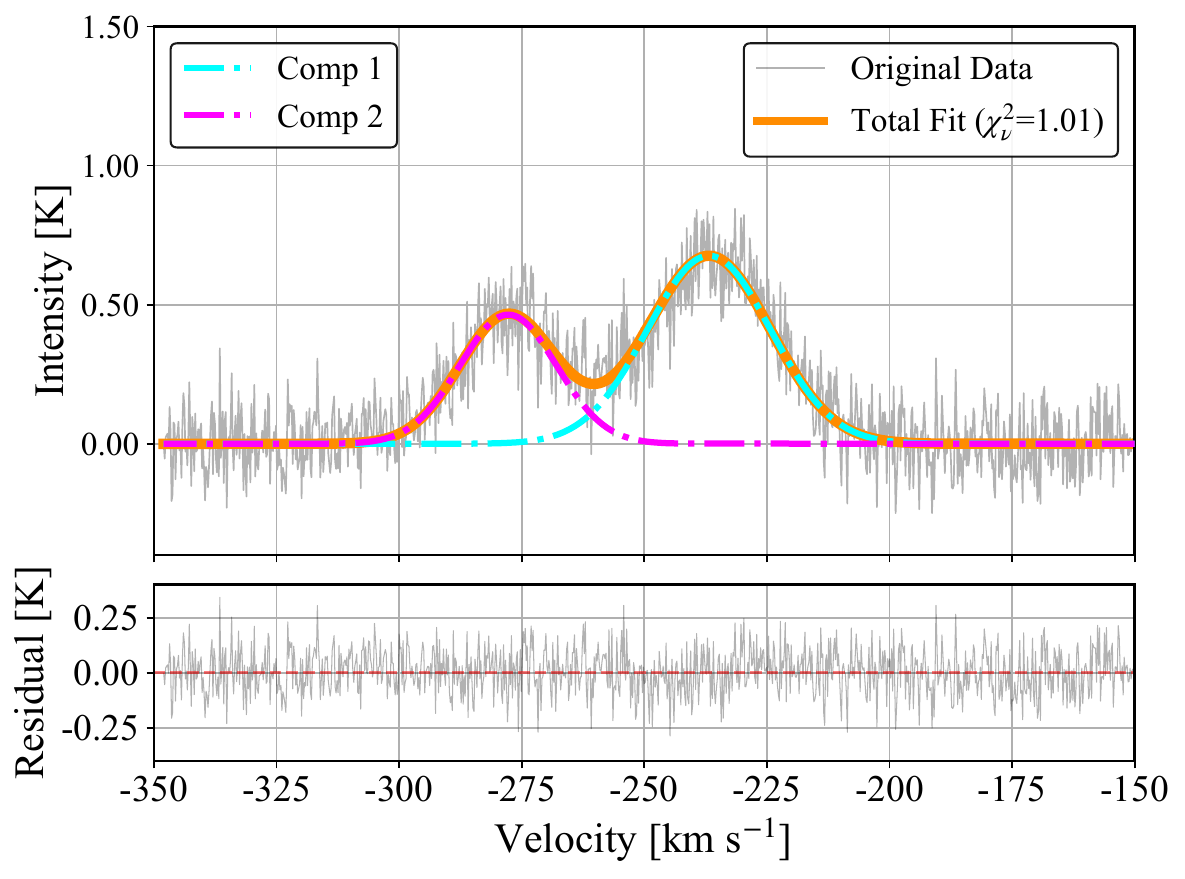}
\label{subfig:spectrum2} 
\end{subfigure}

\vspace{0.3em} 
\begin{subfigure}{0.45\textwidth}
\centering
\includegraphics[width=\linewidth]{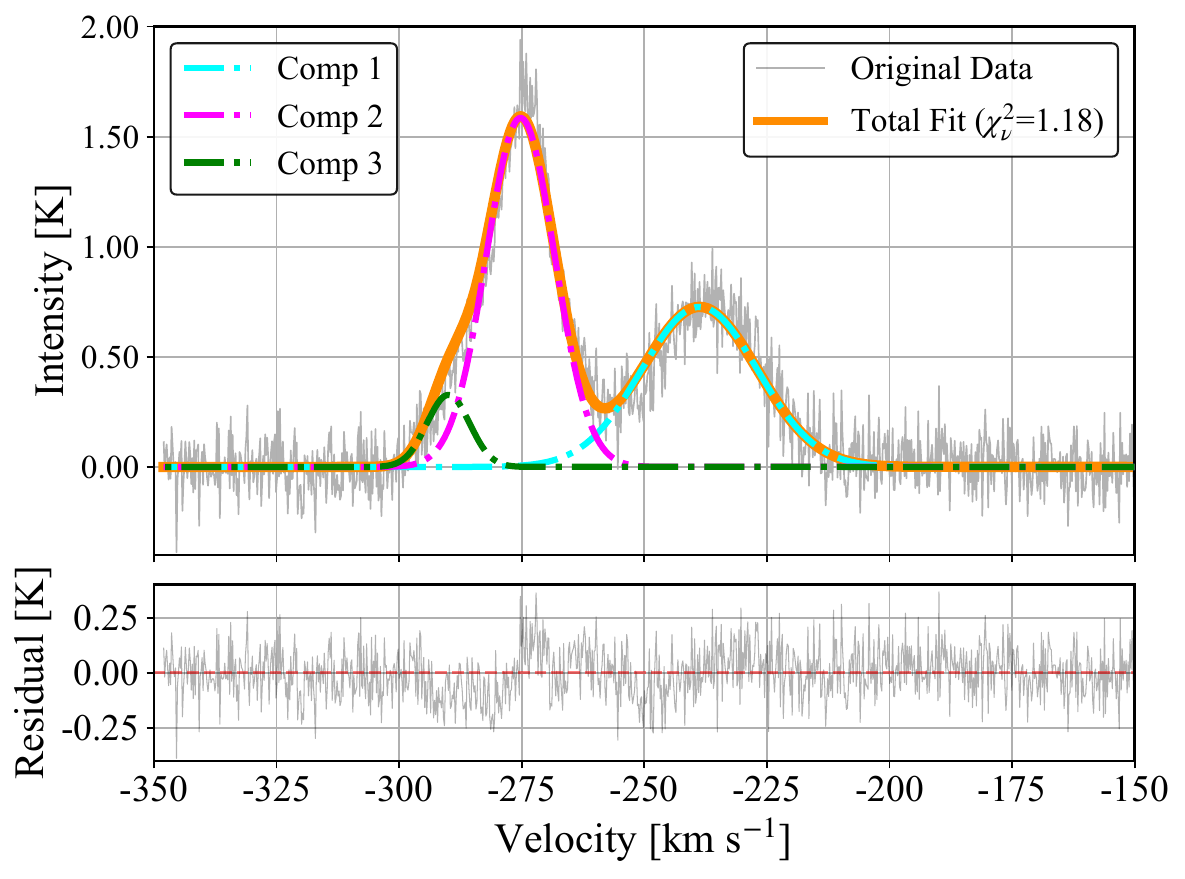}
\label{subfig:spectrum7} 
\end{subfigure}
\hspace{0.02\textwidth} 
\begin{subfigure}{0.45\textwidth}
\centering
\includegraphics[width=\linewidth]{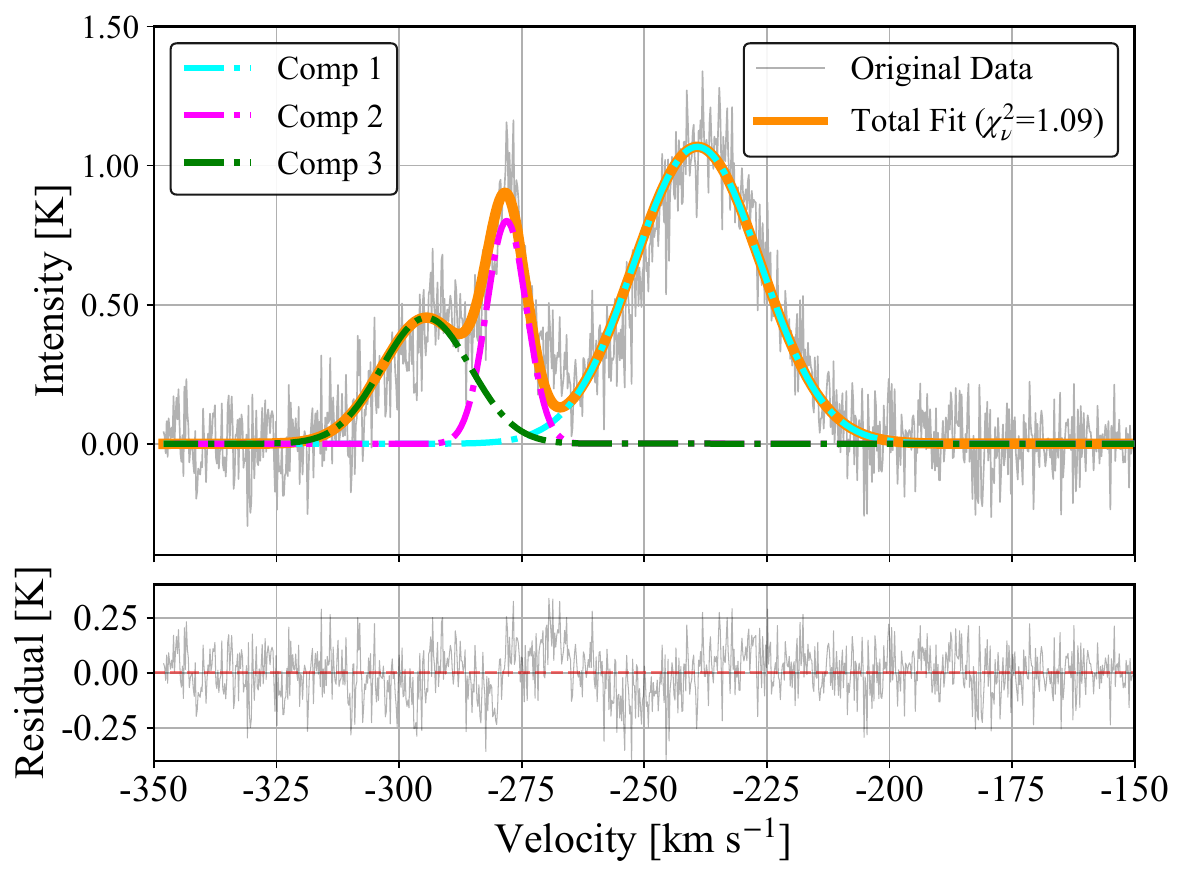} 
\label{subfig:spectrum8}
\end{subfigure}

\caption{Four representative individual-pixel spectra of the cataloged H\,{\sc i} sources, together with the Gaussian fitting results. Only
Gaussian components above Threshold $=2$ are shown. The original spectra are shown in black, with the best-fit in orange and independent Gaussian components in cyan, purple and green. All spectra are fitted using the ROHSA model with $N=3$. In the bottom row, the left panel shows an example that satisfies the adopted merging criteria, whereas the right panel shows an example that does not satisfy the merging criteria.} 
\label{fig:spectrum}
\end{figure*}

 Based on the $N = 3$ decomposition and Threshold = 2, we identified emission features in each Gaussian component after applying the selection criteria. In total, 11 sources were detected across the three components: 6 in the first component, 1 in the second component, and 4 in the third component. Among them, 1 source was identified as a false detection. The basic morphology, spatial distribution, and the number of sources obtained after adopting $\mathrm{FWHM\ Range}=0.7$ and applying the merging procedure are shown in Figure~\ref{fig:CATALOG}. The maps shown in Figure~\ref{fig:CATALOG} are constructed from the ROHSA Gaussian fitting results, rather than directly from the original data cube. Figure~\ref{fig:spectrum} shows the original shapes and best Gaussian fitting results of 4 representative spectra, as well as the residuals, only components above the threshold are shown. 

\subsection{Adopted HI Source Catalog} \label{subsec:catalog}
In this work, the SoFiA-2 mask is used to generate clean Moment 0, Moment 1, and FWHM maps over the full field and to calculate the total integrated flux and H\,{\sc i} mass. The adopted H\,{\sc i} source catalog is constructed from the ROHSA Gaussian component maps using Threshold = 2 and FWHM Range = 0.7. The uncataloged residual component is defined as the difference between the total H\,{\sc i} mass within the SoFiA-masked field and the summed mass of the cataloged H\,{\sc i} sources. Emission outside the SoFiA mask is not included in the scope of this study.

\par Table~\ref{tab:cloud_main} lists the basic parameters derived from the ROHSA Gaussian fitting results for the identified sources. The catalog is arranged with the physical parameters listed in rows and the individual sources shown in Columns~2$-$11, while Column~1 gives the name of each parameter. The catalog includes the centroid RA, the centroid Dec, and the angular area of each source in row 1$-$3; H\,{\sc i} mass in row 4; mean $\mathrm{V}_{\mathrm{LSR}}$ in Row 5; mean $\mathrm{FWHM}$ in row 6; peak $\mathrm{T}_{\mathrm{K,max}}$ and its uncertainty in row 7; integrated flux $\mathrm{F}_{\mathrm{int}}$ and its uncertainty in row 8; peak $\mathrm{T}_{\mathrm{B}}$ in row 9; mean $\mathrm{rms}$ in row 10; and peak H\,{\sc i} column density, $N_{\rm HI}$, and its uncertainty in row 11; the robustness classification based on the threshold tests in row 12. Sources classified as ``R'' (Robust) remain visible when the catalog threshold is increased to Threshold $=3$. ``M'' (Marginal) sources remain visible at Threshold $=2.5$ but not at Threshold $=3$, whereas ``T'' (Threshold-dependent) sources are detected only in the adopted Threshold $=2$ catalog. Source 6 is retained as a merged cataloged source according to the adopted operational criterion. Because it contains two kinematic components, its properties are reported separately as Source 6a and Source 6b in Table~\ref{tab:cloud_main}. The uncertainties for $T_{k,\max}$ are calculated as follows:
\begin{equation}
\begin{aligned}
\mathrm{err}_{T_{\mathrm{K,max}}} &= 21.9 \times 2 \times \mathrm{FWHM} \times \sigma_{\mathrm{FWHM}}. \\
\end{aligned}
\end{equation}
Since the uncertainty of the FWHM is very small, we conservatively take the velocity resolution of the data to be $\sigma_{\rm FWHM}$, $\sigma_{\rm FWHM}\simeq 0.2~{\rm km~s^{-1}}$.
The H\,{\sc i} column density was calculated from the Gaussian-integrated
intensity under the optically thin assumption. For each Gaussian component,
the integrated intensity is
\begin{equation}
W_{\rm HI} = A\sigma_v\sqrt{2\pi},
\end{equation}
where \(A\) is the fitted peak brightness temperature and \(\sigma_v\)
is the velocity dispersion. The corresponding H\,{\sc i} column density is:
\begin{equation}
N_{\rm HI}
=
1.8\times10^{18} W_{\rm HI}
\quad {\rm cm^{-2}} .
\end{equation}

The uncertainty in \(N_{\rm HI}\) was estimated by propagating the
uncertainty in the integrated intensity. The statistical uncertainty in
\(W_{\rm HI}\) is:
\begin{equation}
\sigma_{W,{\rm rms}}
=
\sigma_T \Delta v \sqrt{N_{\rm line}},
\end{equation}
where \(\sigma_T\) is the rms noise per velocity channel, \(\Delta v\)
is the channel width, and \(N_{\rm line}\simeq {\rm FWHM}/\Delta v\)
is the number of channels across the line. We also include a 10\%
systematic uncertainty, consistent with the uncertainty adopted for the
integrated flux. The total uncertainty in \(W_{\rm HI}\) is therefore:
\begin{equation}
\sigma_W
=
\sqrt{
\sigma_{W,{\rm rms}}^2
+
\left(0.1 W_{\rm HI}\right)^2
}.
\end{equation}
The column density uncertainty is then:
\begin{equation}
\mathrm{err}_{N_{\rm HI}}
=
1.8\times10^{18}\sigma_W .
\end{equation}

\par The adopted H\,{\sc i} source catalog accounts for
$M_{\rm cat} \simeq 4.2 \times 10^{6}(d/120\,{\rm kpc})^{2} M_\odot$,
leaving $M_{\rm rem} \simeq 1.1 \times 10^{6}(d/120\,{\rm kpc})^{2} M_\odot$,
or about 20\%, outside the cataloged sources.
Among the 9 identified sources, Source 1 alone contributes about 93\% of the cataloged H\,{\sc i} mass and about 74\% of the total H\,{\sc i} mass in this field.
This suggests that this region is better described as a dominant H\,{\sc i} complex accompanied by several smaller kinematic components.

\begingroup
\setlength{\tabcolsep}{2pt} 
\renewcommand{\arraystretch}{1}
\begin{deluxetable*}{lcccccccccc}
\tabletypesize{\scriptsize}
\tablewidth{0pt}
\tablecaption{HVC Cloud Parameters \label{tab:cloud_main}}
\tablehead{
\colhead{Parameter} & \colhead{Source~1} & \colhead{Source~2} & \colhead{Source~3} & \colhead{Source~4} & \colhead{Source~5} & \colhead{Source~6a} & \colhead{Source~6b} & \colhead{Source~7} & \colhead{Source~8} & \colhead{Source~9}
}
\startdata
RA ($^{\circ}$)
& $355.5$ & $353.9$ & $353.7$ & $353.6$ & $353.7$
& $356.2$ & $356.2$ & $356.2$ & $356.1$ & $356.4$ \\
Dec ($^{\circ}$)
& $-5.9$ & $-6.7$ & $-6.4$ & $-5.5$ & $-5.1$
& $-6.6$ & $-6.6$ & $-6.5$ & $-6.3$ & $-4.9$ \\
Angular area (arcmin$^2$) & $11801.3$ & $321.8$ & $38.3$ & $243.0$ & $49.5$ & $677.3$ & $677.3$ & $85.5$ & $24.8$ & $42.8$\\
$M_{\rm H\,I}^{a}$
[$10^{3}(d/120\,{\rm kpc})^{2}M_\odot$]
& $3931.2 \pm 393.4$
& $83.5 \pm 8.1$
& $8.1 \pm 0.6$
& $46.7 \pm 4.6$
& $8.1 \pm 0.6$
& $152.1 \pm 15.0$
& $2.9 \pm 0.6$
& $10.4 \pm 1.2$
& $2.3 \pm 0.2$
& $4.6 \pm 0.6$ \\  
$V_{\rm LSR}$ (km s$^{-1}$) & $-238.4$ & $-235.4$ & $-233.2$ & $-242.9$ & $-234.3$ & $-276.7$ & $-290.7$ & $-294.9$ & $-292.5$ & $-293.4$  \\
FWHM (km s$^{-1}$) & $31.9$ & $46.9$ & $46.7$ & $39.4$ & $36.8$ & $20.1$ & $14.1$ & $19.6$ & $21.6$ & $21.6$  \\
${T_{K,\max}}^{b}$ ($10^3$ K) & $22.2 \pm 0.3$ & $48.1 \pm 0.4$ & $47.7 \pm 0.4$ & $34.0 \pm 0.3$ & $29.7 \pm 0.3$ & $8.9 \pm 0.2$ & $4.3 \pm 0.1$ & $8.4 \pm 0.2$ & $10.2 \pm 0.2$ & $10.2 \pm 0.2$  \\
$F_{\rm int}$ (Jy km s$^{-1}$) & $1158.8 \pm 115.9$ & $24.6 \pm 2.5$ & $2.4 \pm 0.2$ & $13.7 \pm 1.4$ & $2.4 \pm 0.2$ & $44.8 \pm 4.5$ & $0.9 \pm 0.1$ & $3.1 \pm 0.3$ & $0.7 \pm 0.1$ & $1.4 \pm 0.1$  \\
Peak $T_B$ (K) & $1.6$ & $0.4$ & $0.3$ & $0.3$ & $0.3$ & $2.0$ & $0.4$ & $0.5$ & $0.3$ & $0.3$  \\
rms (K) & $0.1$ & $0.1$ & $0.1$ & $0.1$ & $0.1$ & $0.1$ & $0.1$ & $0.1$ & $0.1$ & $0.1$ \\
$N_{\rm HI,max}$ ($10^{19}$ cm$^{-2}$) & $8.5 \pm 0.8$ & $3.6 \pm 0.4$ & $2.4 \pm 0.3$ & $2.4 \pm 0.2$ & $2.0 \pm 0.2$ & $5.7 \pm 0.6$ & $0.9 \pm 0.1$ & $2.0 \pm 0.2$ & $1.2 \pm 0.1$ & $1.5 \pm 0.2$  \\
Flag$^{c}$ & R & R & T & M & T& R & T & R & T & M \\
\enddata
\tablecomments{$^{a}$ Uncertainties take into account statistical and systematic errors, see Equation 6.\\
$^{b}$The linewidth-derived temperatures are listed as upper limits, $T_{k,\max}$, under the assumption of purely thermal broadening. \\
$^{c}$Row 12 lists the robustness classification based on the threshold tests. ``R'' (Robust) denotes sources that are still recovered at Threshold $=3$; ``M'' (Marginal) denotes sources recovered at Threshold $=2.5$ but not at Threshold $=3$; and ``T'' (Threshold-dependent) denotes sources recovered only in the adopted Threshold $=2$ catalog. Source 6 is split into Source 6a and Source 6b, corresponding to two kinematic components of the merged structure.
}
\end{deluxetable*}
\endgroup

\subsection{Kinematics \& projection test} \label{subsec:P-V}
\begin{figure*}[ht!]
\centering

\begin{subfigure}{0.49\textwidth}
\centering
\includegraphics[width=\linewidth]{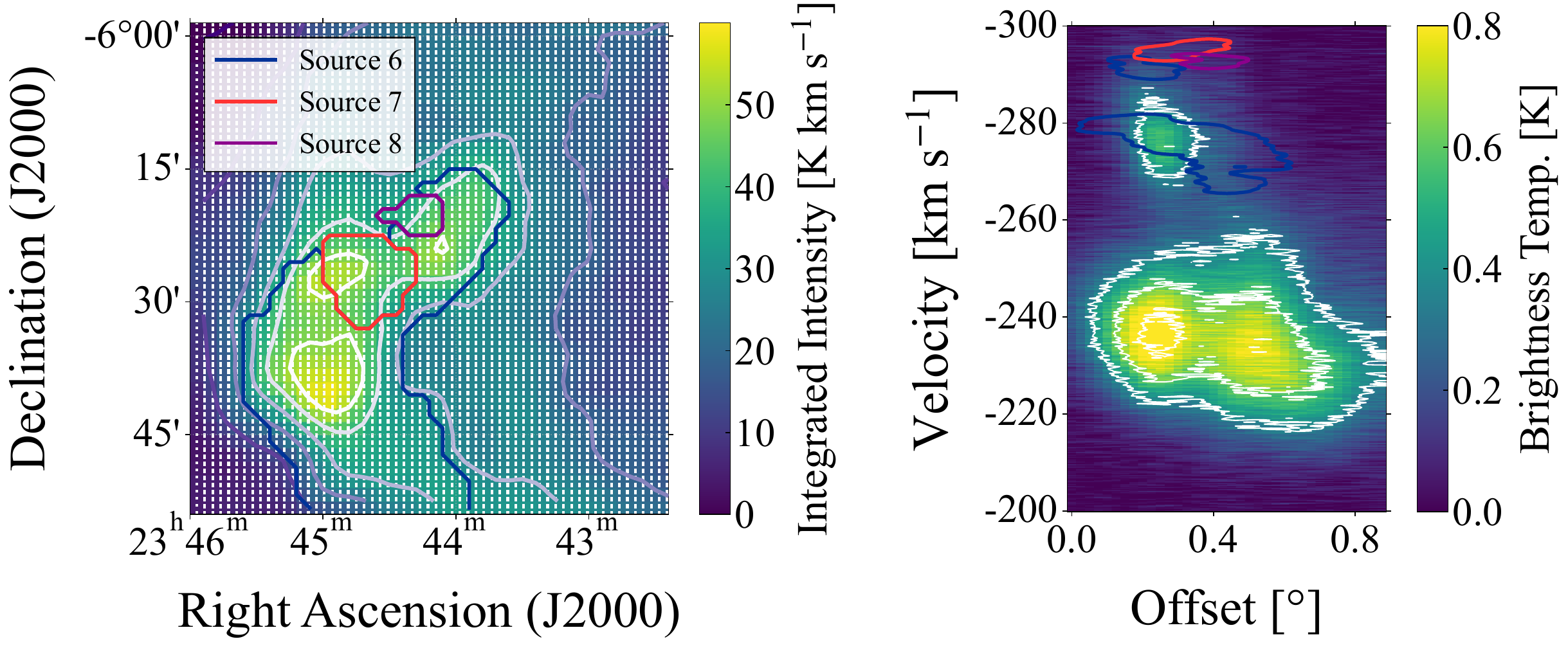}
\caption{0$^\circ$}
\label{subfig:15}
\end{subfigure}
\hfill
\begin{subfigure}{0.49\textwidth}
\centering
\includegraphics[width=\linewidth]{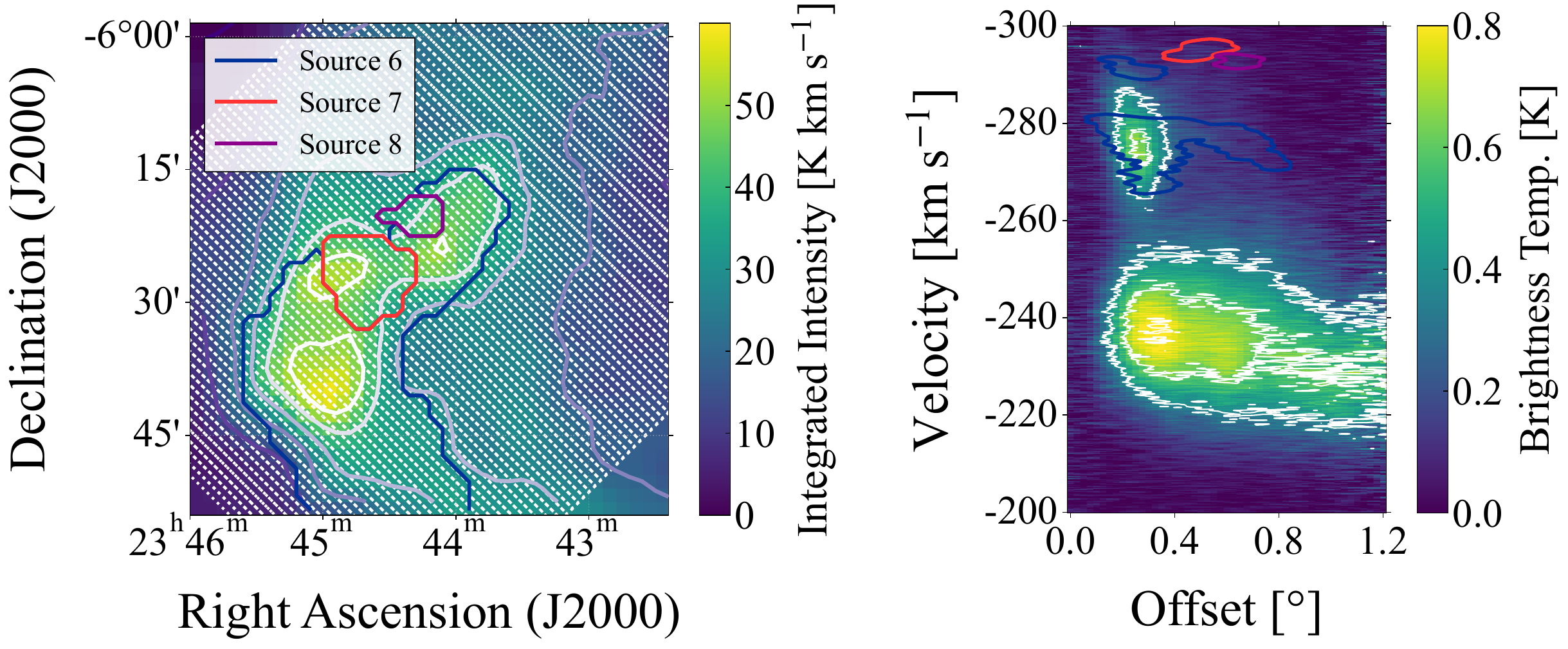}
\caption{45$^\circ$}
\label{subfig:30}
\end{subfigure}

\vspace{1em}

\begin{subfigure}{0.49\textwidth}
\centering
\includegraphics[width=\linewidth]{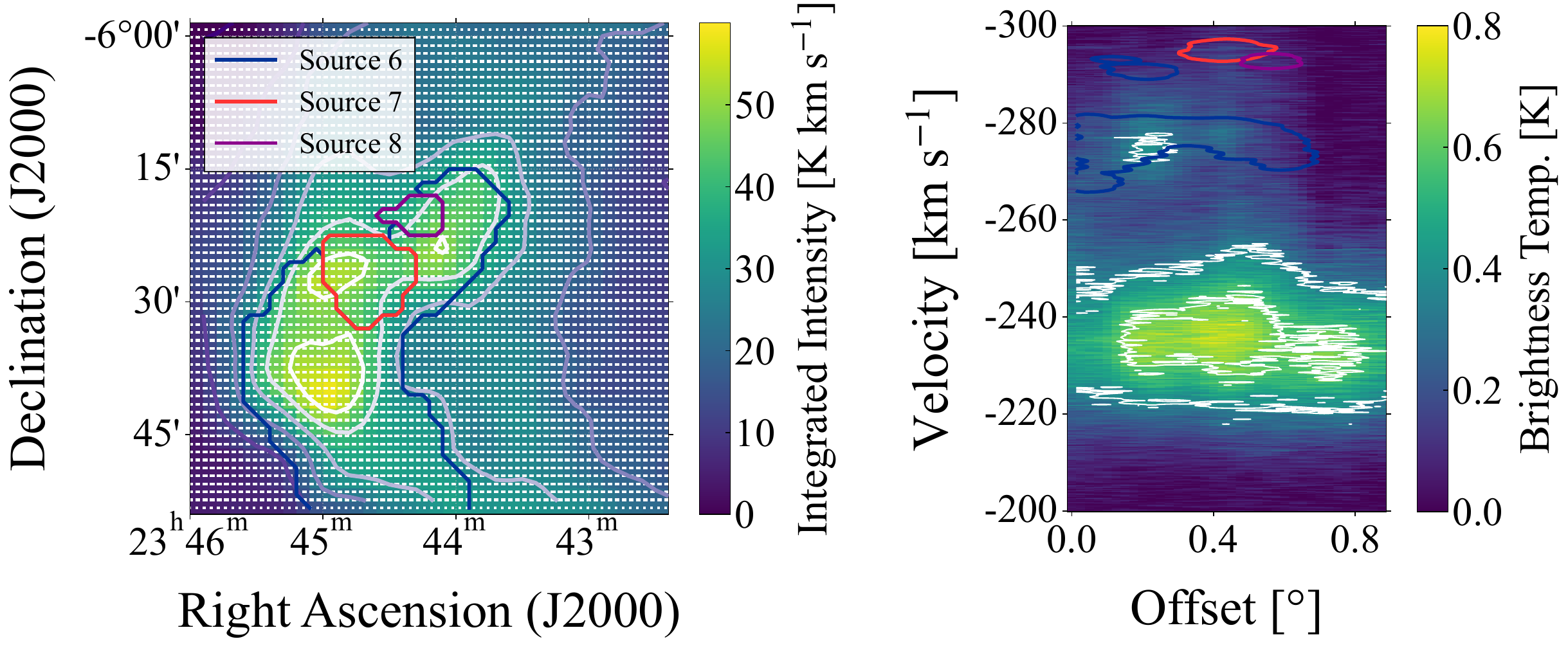}
\caption{90$^\circ$}
\label{subfig:45}
\end{subfigure}
\hfill
\begin{subfigure}{0.49\textwidth}
\centering
\includegraphics[width=\linewidth]{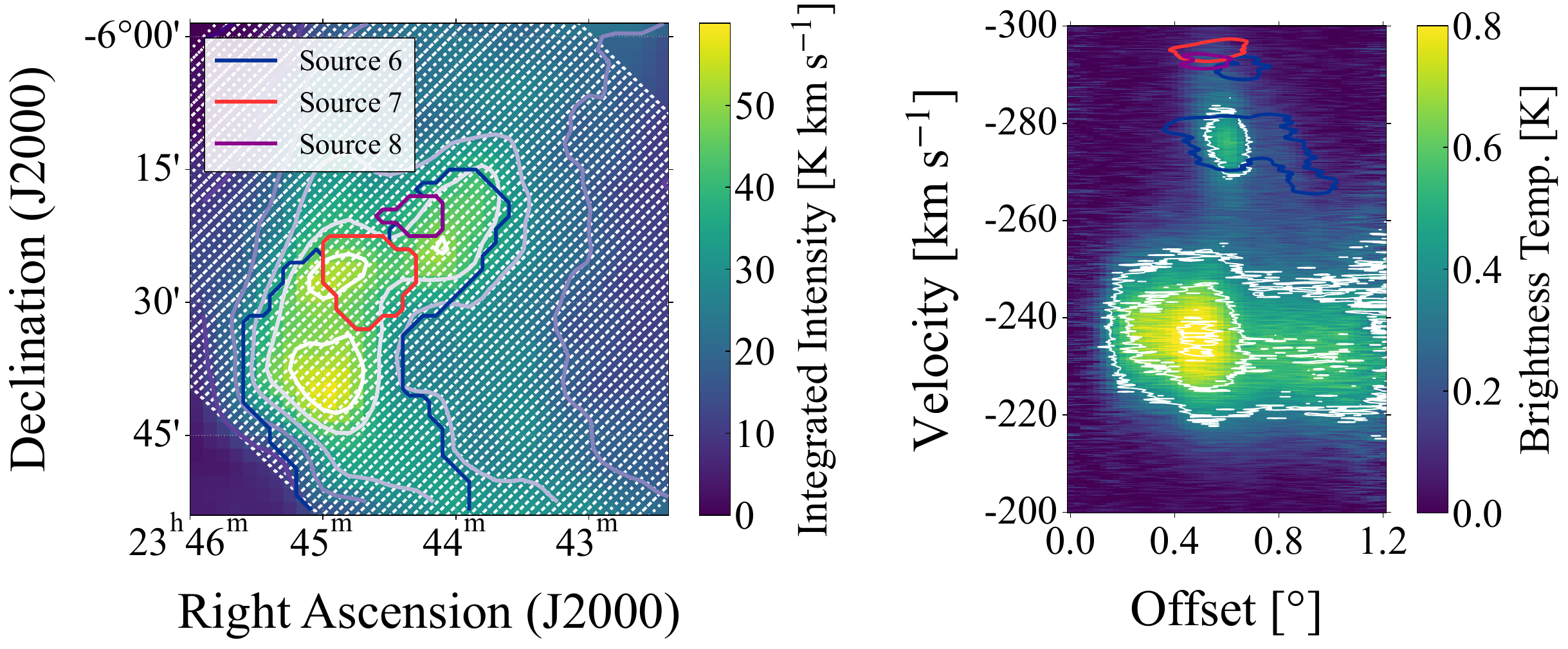}
\caption{135$^\circ$}
\label{subfig:30-0}
\end{subfigure}

\caption{Integration paths with different position angles (from 0$^\circ$ to 135$^\circ$) are constructed using the original data. In each panel, the left image shows the integration path and width, indicated by white points, with an integration width of 60 arcmin. For better visualization, the colorbar ranges from 0 K km s$^{-1}$ to 60 K km s$^{-1}$, while the overlaid contours follow the same contour levels as in Figure~\ref{fig:CRAFTS_HI4PI_moments}. The blue, red, and purple contours outline Source 6, Source 7, and Source 8, respectively, with Source 6 further divided into Sources 6a and 6b. Source 1 extends over nearly the entire displayed area. The right image presents the corresponding P-V diagram, overlaid with 3, 5, and 7$\sigma$ contours at 0.36 K, 0.60 K, and 0.84 K. Source boundaries are marked using the same colors as in the left image.}
\label{fig:all-P-V}
\end{figure*}

Sources 6, 7, 8, and Source 1 appear spatially connected in the Moment 0, Moment 1, and FWHM maps, with a velocity difference of approximately 40 $-$ 60 km s$^{-1}$. To test whether the apparent spatial overlap among Sources 6, 7, 8, and the extended Source 1 is associated with physical interaction, we extracted P-V diagrams along several representative directions across the overlap region, as shown in Figure~\ref{fig:all-P-V}. The P-V diagrams were constructed from the unmasked CRAFTS H\,{\sc i} data cube, using extraction paths with position angles from 0$^\circ$ to 135$^\circ$ and a path width of $60~{\rm arcmin}$. For each direction, we also extracted the subregion containing Sources 6, 7, and 8, and overlaid their source boundaries on both the spatial maps and the corresponding P-V diagrams. Source 6 is further separated into Sources 6a and 6b following its resolved morphology. The resulting P-V diagrams do not show clear spatial-kinematic continuity or intermediate-velocity bridge emission between the velocity components. 

\subsection{Thermal limits from line width} \label{subsec:Thermal}
\par To constrain the kinetic temperature of the H\,{\sc i} gas, we use the measured linewidth as a kinematic proxy. For comparison, the typical velocity dispersion of HVCs derived from the LAB survey is 
$\sigma \sim 11$$-$$13\,{\rm km\,s^{-1}}$, corresponding to 
${\rm FWHM} \sim 25$$-$$30\,{\rm km\,s^{-1}}$ \citep{2006A&A...455..481K}. If the observed linewidth is assumed to arise purely from thermal broadening, it gives an upper limit on the kinetic temperature,
\begin{equation}
T_{\rm k,max}
=
\frac{m_{\rm H}}{8 k_{\rm B}\ln 2}
\,{\rm FWHM}^{2}
=
21.9\,{\rm FWHM}^{2},
\end{equation}
where FWHM is in km s$^{-1}$, $m_{\rm H}$ is the hydrogen atom mass, and $k_{\rm B}$ is the Boltzmann constant \citep{2011piim.book.....D}. 

According to Table~\ref{tab:cloud_main}, the FWHM values of the nine cataloged H\,{\sc i} sources range from 14.1 to 46.9 km s$^{-1}$, corresponding to 
$T_{\rm k,max} \sim 4.3\times10^{3}$$-$$4.8\times10^{4}$ K under the purely thermal assumption. These values should be regarded as upper limits, because turbulence, unresolved velocity gradients, and blended velocity components may also contribute to the observed linewidths.

\section{DISCUSSION} \label{sec:discussion}
\subsection{Projection versus direct collision} \label{subsec:Fragmentation}

Although clearly separated velocity components are present, the P-V diagram does not reveal a connecting bridge feature between them. We also tested alternative P-V extraction directions, but did not find continuous intermediate-velocity emission linking the velocity components. In the cloud-cloud collision scenario, colliding clouds may show characteristic observational signatures, including an intermediate-velocity bridge connecting the two velocity components, a continuous or V-shaped structure in P-V space, and complementary spatial distributions between the interacting clouds \citep{2021PASJ...73S...1F}. In our data, neither a clear bridge feature nor a V-shaped P-V structure is detected. This suggests that the apparent overlap is more likely explained by line-of-sight projection than by a direct cloud-cloud collision. Therefore, we find no clear P-V evidence for a direct cloud-cloud collision in this field, although weak interactions below our sensitivity or resolution limits cannot be fully excluded.

\subsection{Relation to global MS filamentary structure} \label{subsec:multi-phase}
\par Previous studies have suggested that the MS may consist of two extended filaments originating from the LMC and the SMC, respectively. Figure 1 of \citet{2015ApJ...813..110H} shows the $N_{\rm HI}$ map of the MS over the range $-150^\circ < L_{\rm MS} < 70^\circ$ and $-40^\circ < B_{\rm MS} < 40^\circ$, indicating the overall position and morphology of the MS. The figure also shows two filamentary structures traced by dashed lines in different colors, representing material originating from the LMC and the SMC, respectively. Based on this interpretation, the region studied in this work may be associated with material originating from the SMC. A definitive confirmation would require measurements based on metal absorption lines, which are beyond the scope of this study. More recently, \citet{2025OJAp....8E..16Z} separated the MS into a dominant H\,{\sc i} strand and a sub-dominant strand associated with a stellar population that may originate from the SMC. Our field ($-84.9^\circ \lesssim L_{\rm MS}\lesssim-81.6^\circ$ and $-3.4^\circ\lesssim B_{\rm MS}\lesssim1^\circ$) overlaps in projection with their dominant rather than sub-dominant strand. In addition, one Magellanic Stellar Stream candidate identified by \citet{2023ApJ...956..110C}, located at $\mathrm{RA}=353.1^\circ$ and $\mathrm{Dec}=-5.0^\circ$, lies within our observed field. This star is at a distance of $107\pm5$ kpc, but its approximate LSR velocity of $-187\ {\rm km\,s^{-1}}$ differs by about $50\ {\rm km\,s^{-1}}$ from the dominant CRAFTS component at $-238.4\ {\rm km\,s^{-1}}$. Thus, although a projected connection to the distant stellar structure is possible, the present data do not establish a physical association. Gas metallicity and distance measurements are required to distinguish between an LMC, SMC, or mixed origin. The multi-filamentary morphology observed in our field is broadly consistent with the filamentary and fragmented structure of the MS, which is known to comprise multiple coherent filaments and clumpy small-scale structures\citep{2010ApJ...723.1618N, 2003ApJ...586..170P}. 

\par The GALFA observations targeted the extreme tip of the MS, farther downstream from our MS IV field, and resolved four narrow H\,{\sc i} filaments, S1$-$S4 \citep{2008ApJ...680..276S}. Along S1, the velocity changes by about $75\,{\rm km\,s^{-1}}$ over $10^\circ$. Extrapolating this trend toward our field gives $V_{\rm LSR}\sim-225\,{\rm km\,s^{-1}}$, close to the dominant CRAFTS component at approximately $-240\,{\rm km\,s^{-1}}$. However, the lack of contiguous high-resolution H\,{\sc i} coverage prevents us from establishing morphological continuity or a physical association.

\subsection{Context for future multiphase work} \label{subsec:implications}
\par The MS has long been discussed as a potential source of gas accretion onto the Milky Way, but whether its neutral gas can survive interaction with the hot halo and eventually reach the Galactic disk remains uncertain \citep{2012ARA&A..50..491P,2014ApJ...787..147F}. The survival of  cool/neutral gas depends on processes such as hydrodynamic disruption, turbulent mixing, radiative cooling, and condensation of hot halo gas in cloud wakes \citep{2017MNRAS.470..114A,2018MNRAS.480L.111G}. These processes are inherently multiphase and cannot be constrained by H\,{\sc i} emission alone.

\par Our FAST observations provide a localized view of the neutral component in one MS IV field. The coherent filamentary H\,{\sc i} structures and smaller kinematic components show that neutral gas in this region remains structured on arcminute scales. However, the present H\,{\sc i} data do not determine whether these structures will survive in the halo or eventually be accreted onto the Galactic disk. Addressing their long-term evolution will require numerical simulations and multiphase observations, especially UV absorption-line measurements of the associated ionized gas.

\section{Systematics and Robustness} \label{sec:Systematics}
In this section, we summarize the main systematic uncertainties and robustness tests relevant to the flux measurements, source identification, and physical interpretation. 
The dominant uncertainties arise from baseline correction, flux calibration, the adopted mask and resolution for the CRAFTS-HI4PI comparison, and the parameter dependence of the ROHSA-based source catalog.

As a control test, we applied the same SoFiA-2 masking and source-finding procedure to the signal-free velocity range of $-400~{\rm km~s^{-1}} \lesssim V_{\rm LSR} \lesssim -350~{\rm km~s^{-1}}$. No valid sources were detected in this control region, indicating that the adopted mask and source-finding parameters do not generate significant false detections in pure-noise channels.

We also tested the robustness of the ROHSA-based source identification. As shown in Table~\ref{tab:source_merging_parameters}, the number of sources remaining after removing false detections and applying the merging criterion varies from 5 to 10, depending on the adopted signal-to-noise threshold and FWHM Range. Therefore, we aim to identify and characterize emission as completely as possible down to the sensitivity limit, and the exact number of sources is not considered a primary result of this work. The fiducial choice of Threshold = 2 and FWHM Range = 0.7 yields nine sources, which we use as a working catalog for measuring fluxes, masses, line widths, and column densities. 
Sources that are still recovered when the threshold is increased to 
Threshold = 3 are classified as Robust (R). Sources that remain recovered 
at Threshold = 2.5 but disappear at Threshold = 3 are classified as 
Marginal (M). Sources detected only in the adopted Threshold = 2 catalog 
are classified as Threshold-dependent (T).

The mass budget is also dominated by Source 1. Thus, the catalog should be interpreted as one dominant filamentary H\,{\sc i} complex plus several smaller kinematic components, rather than as a collection of many equally important clumps. This interpretation is consistent with the parameter tests: the large-scale filaments and the dominant source remain stable, while the faintest sources depend on the adopted threshold and merging criterion.

\section{Conclusions} \label{sec:Conclusions}

\par We present CRAFTS H\,{\sc i} observations of a $3.8^\circ\times2.2^\circ$ field in the MS IV region. The high-sensitivity FAST data resolve the emission into three coherent filamentary H\,{\sc i} structures, rather than a single diffuse cloud. Compared with the HI4PI data, the CRAFTS observations reveal finer filamentary and clumpy structure while preserving the same large-scale morphology. The total H\,{\sc i} mass in the analyzed field is $\simeq 5.3 \times 10^{6} (d/120\,{\rm kpc})^{2}\,M_{\odot}$,
where $d$ is the unknown distance to the MS IV gas. 

\par Using ROHSA Gaussian decomposition together with a source-identification and merging procedure, we construct an adopted H\,{\sc i} source catalog for this field. The adopted catalog contains 9 H\,{\sc i} sources for Threshold $=2$ and FWHM Range $=0.7$. Source 1 dominates the mass budget, contributing over 90\% of the cataloged mass and over 70\% of the total field mass. This indicates that the field is dominated by one major filamentary H\,{\sc i} complex, together with several smaller kinematic components.

\par Because Sources 6, 7, and 8 appear spatially overlapped with the extended Source 1 in projection, we further examine their apparent overlap regions using P-V diagrams. The P-V diagrams across the apparent overlap region show no clear intermediate-velocity bridge or V-shaped structure. The available P-V diagrams therefore favor a line-of-sight projection interpretation over a direct cloud-cloud collision scenario, although weak interactions below the present sensitivity or resolution limits cannot be excluded.

\par Overall, the FAST data provide a local, high-sensitivity single-dish view of filamentary and compact H\,{\sc i} structure in this MS IV field. This localized study provides a pilot for applying the same analysis to a wider portion of the MS tail. Wider contiguous H\,{\sc i} mapping with future CRAFTS releases and complementary ASKAP/GASKAP observations will be needed to determine whether the dominant complex remains coherent upstream and downstream of the present field and how it connects to the MS tip. Numerical simulations and multiphase observations, especially UV absorption-line measurements of the associated ionized gas, will further help connect these neutral structures to the broader evolution of the MS.

\begin{acknowledgments}
This work is supported by the National SKA Program of China
No. 2025SKA0150103, and National Natural Science Foundation
of China under Nos. 12550002, 12133008, 12221003, and 11890692.
We acknowledge the science research grants from the China
Manned Space Project with No. CMS-CSST-2021-A04 and
No. CMS-CSST-2025-A10. This work is also supported by the
Fundamental Research Funds for the Central Universities.
Q.Y. is supported by the European Research Council (ERC)
under grant agreement No. 101040751. This work uses the data
from Five-hundred-meter Aperture Spherical radio Telescope
(FAST). FAST is a Chinese national megascience facility,
operated by the National Astronomical Observatories of Chinese
Academy of Sciences.
\end{acknowledgments}

\bibliography{sample701}
\bibliographystyle{aasjournal}

\appendix

\section{Moment maps for Individual Sources}
\label{app:individual_moment_maps}

Here we present the individual Moment 0, Moment 1, and FWHM maps of the nine cataloged sources obtained with Threshold = 2 and FWHM Range = 0.7. These maps are constructed from the ROHSA Gaussian-component results, using the same procedure as in Figure~\ref{fig:CATALOG}. For each source, the panels show the full-field Moment 0 map of the source, the zoomed-in Moment 0 map, the Moment 1 map, and the Gaussian-fit FWHM map.

\begin{figure*}[ht!]
\centering
\includegraphics[width=0.95\textwidth]
{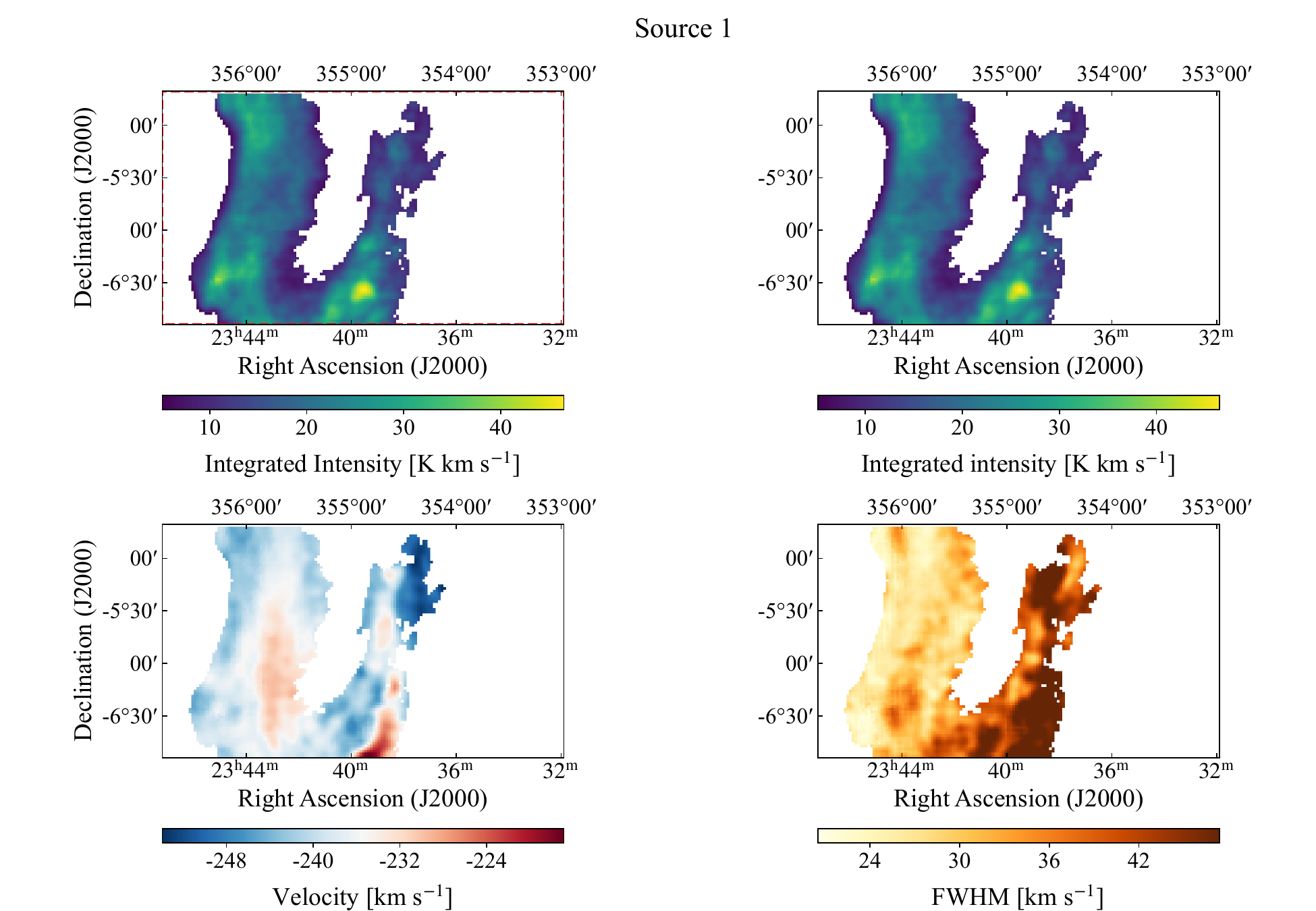}
\caption{Moment maps for Sources 1--9. For each source, the first panel in the first row shows the Moment 0 map over the full field. The red dashed box marks the zoom-in region shown in the remaining three panels. The other three panels show the zoomed-in Moment 0, Moment 1, and FWHM maps, respectively.}
\label{fig:CATALOG_all}
\end{figure*}
\begin{figure*}[!t]
\centering
\includegraphics[width=0.95\textwidth]
{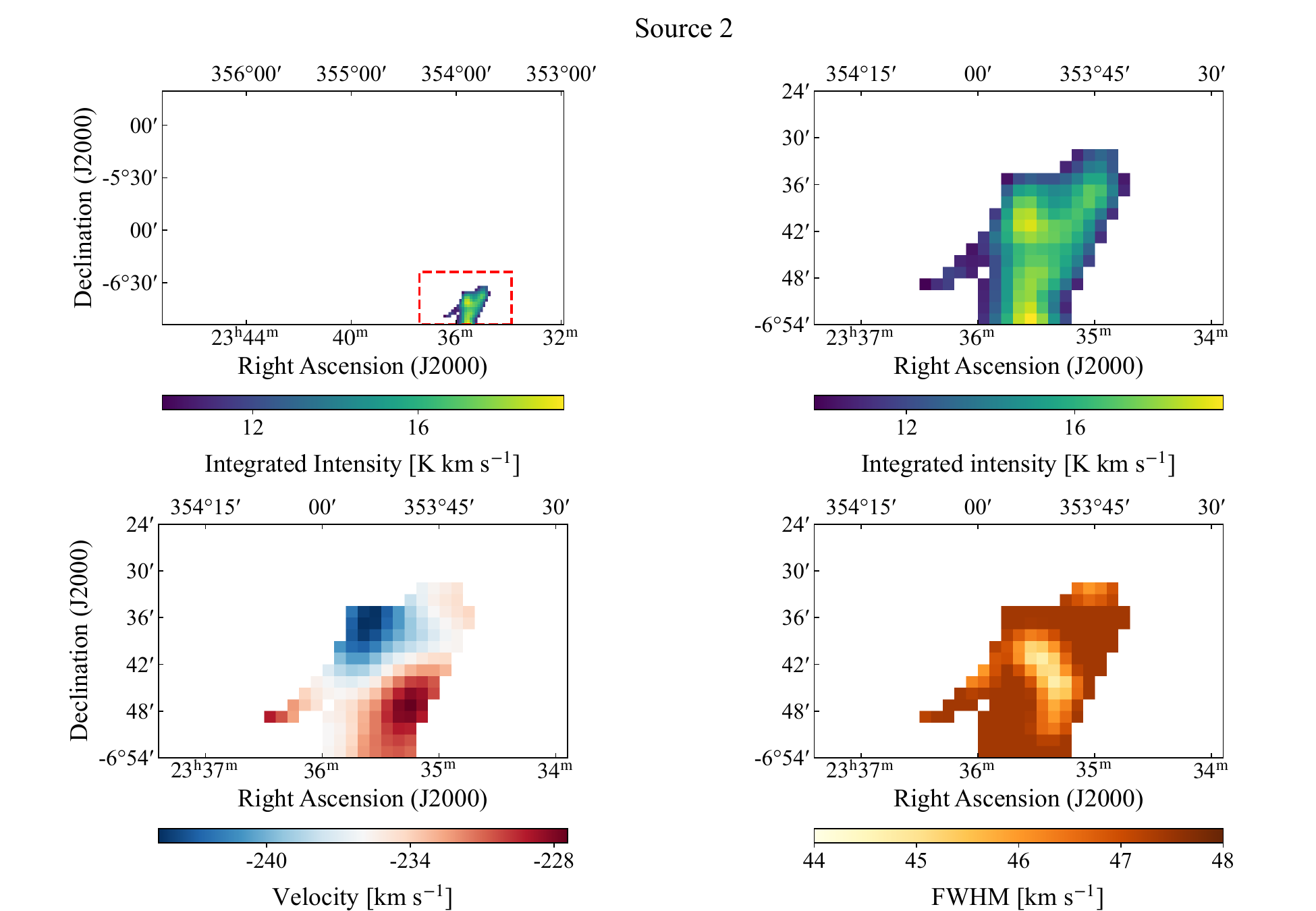}
\vspace{0.3em}
\includegraphics[width=0.95\textwidth]
{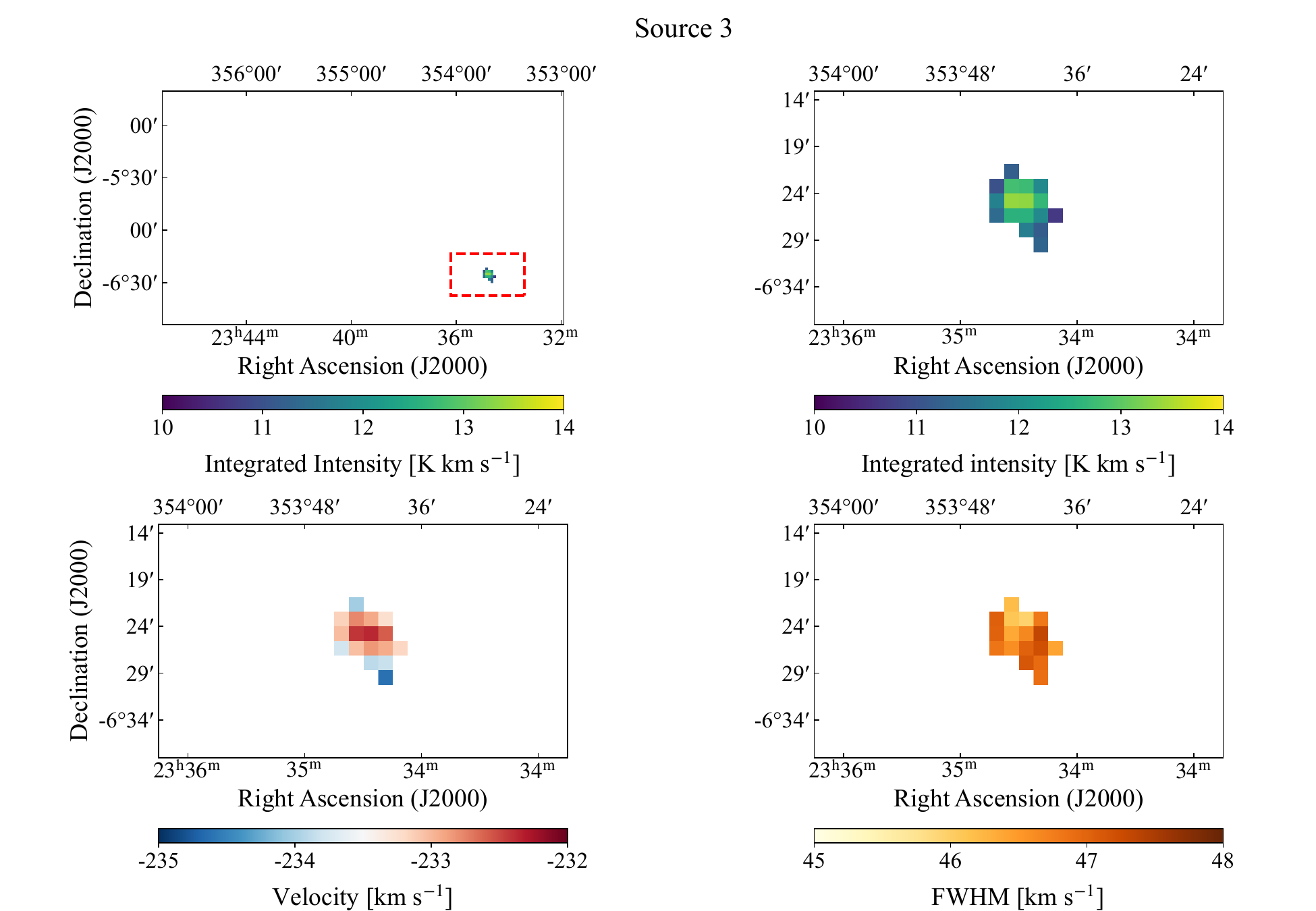}
\addtocounter{figure}{-1}
\caption{Continued.}
\end{figure*}
\begin{figure*}[!t]
\centering
\includegraphics[width=0.95\textwidth]
{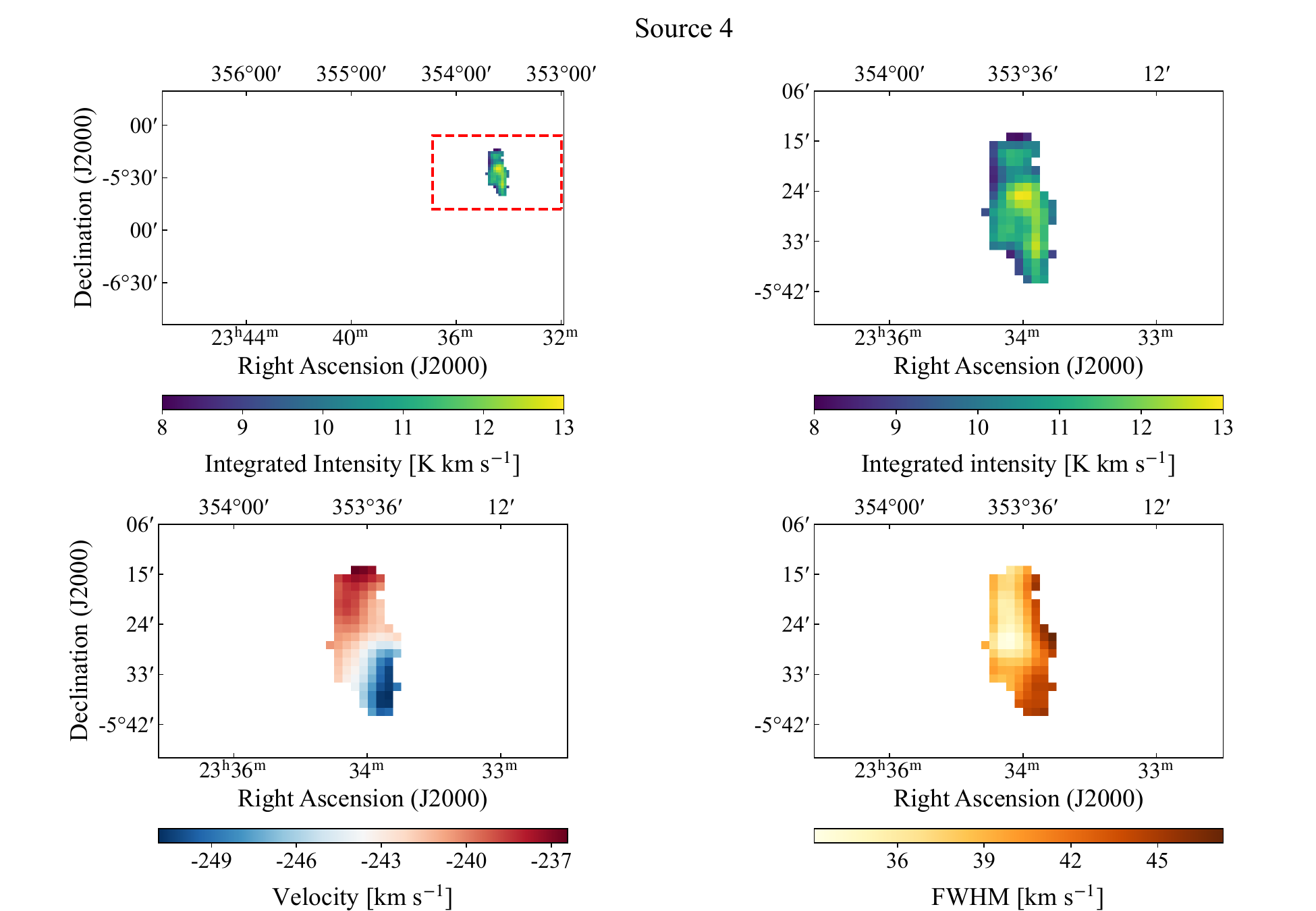}
\vspace{0.3em}
\includegraphics[width=0.95\textwidth]
{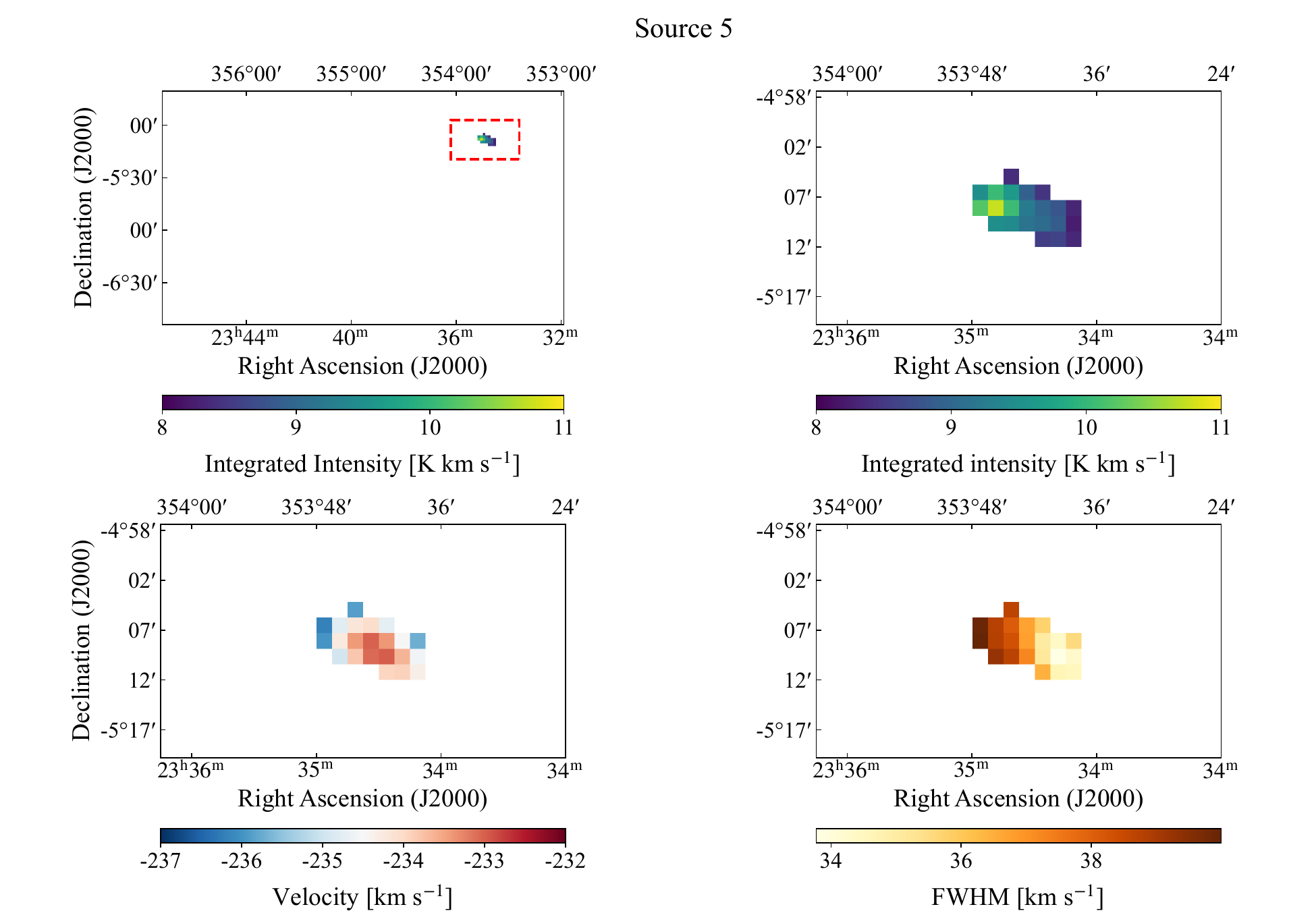}
\addtocounter{figure}{-1}
\caption{Continued.}
\end{figure*}
\begin{figure*}[!t]
\centering
\includegraphics[width=0.95\textwidth]
{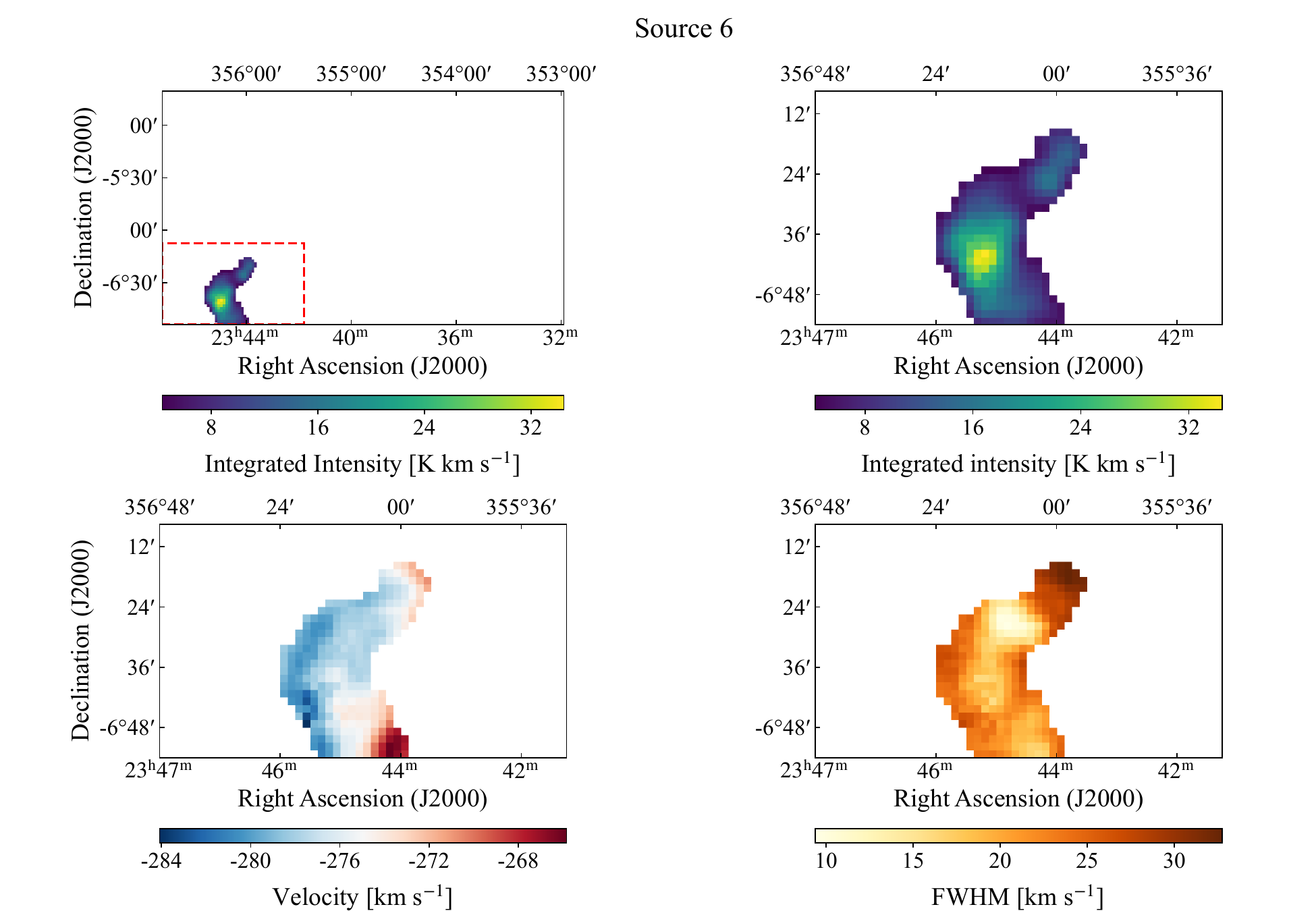}
\vspace{0.3em}
\includegraphics[width=0.95\textwidth]
{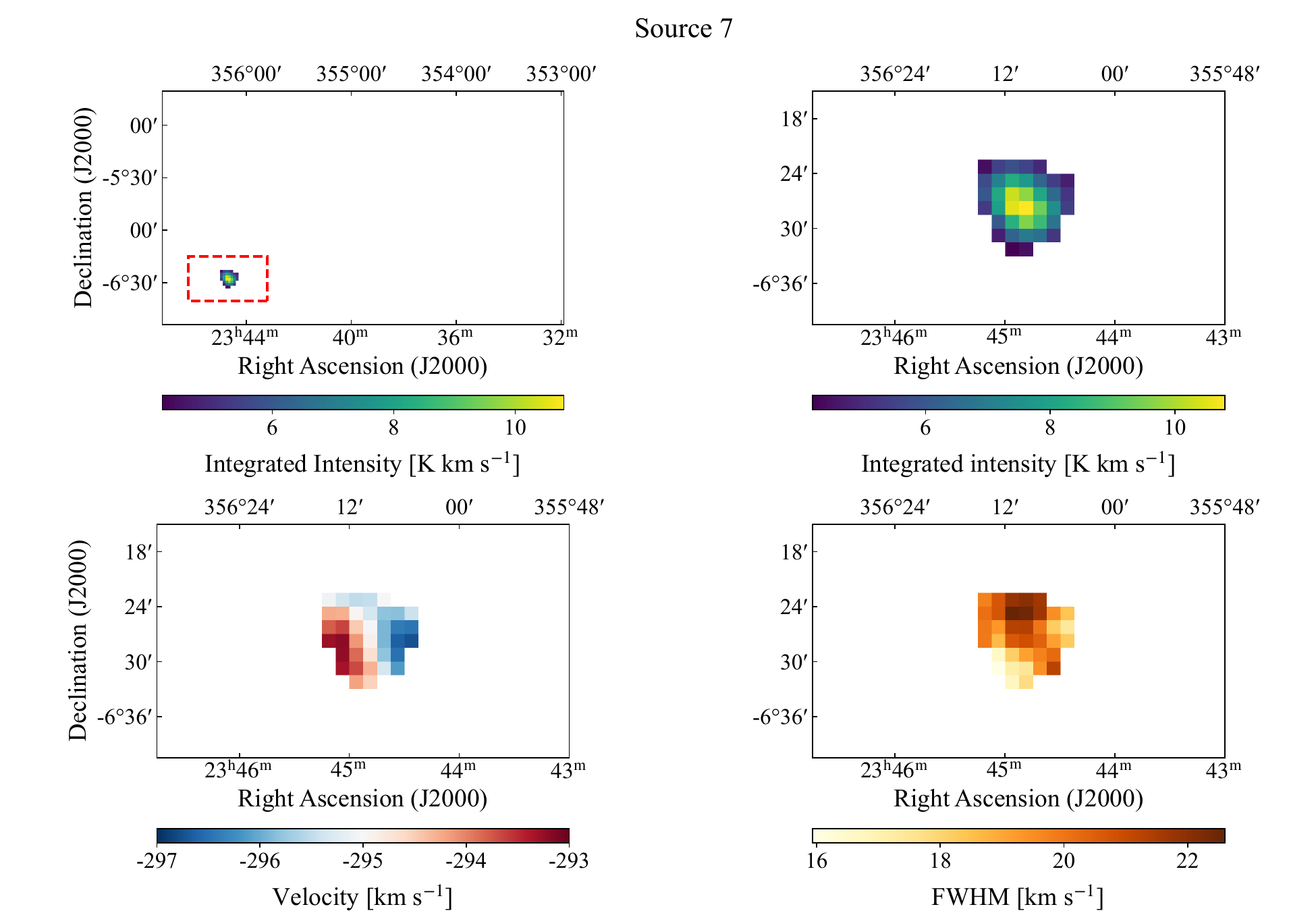}
\addtocounter{figure}{-1}
\caption{Continued.}
\end{figure*}
\begin{figure*}[!t]
\centering
\includegraphics[width=0.95\textwidth]
{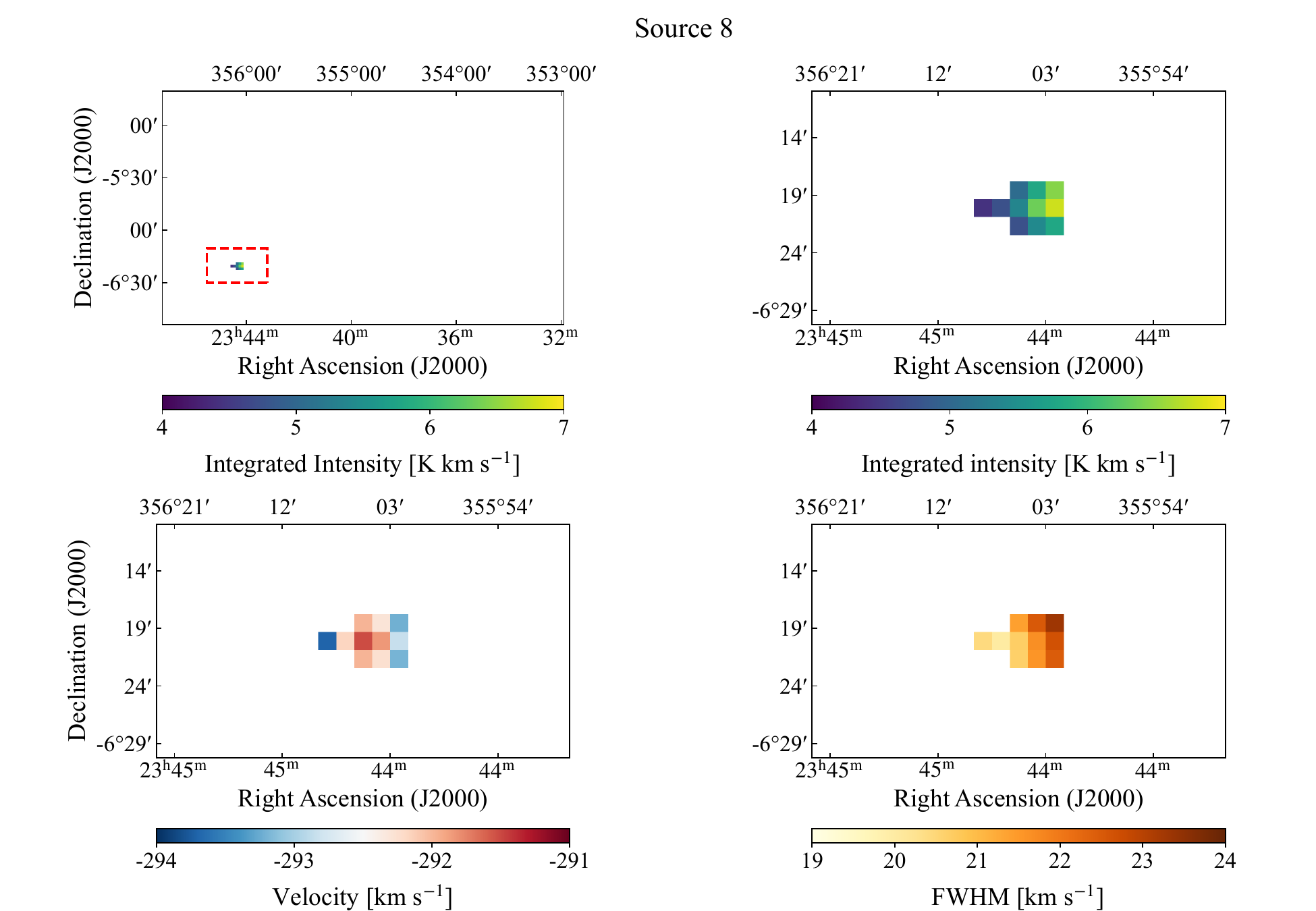}
\vspace{0.3em}
\includegraphics[width=0.95\textwidth]
{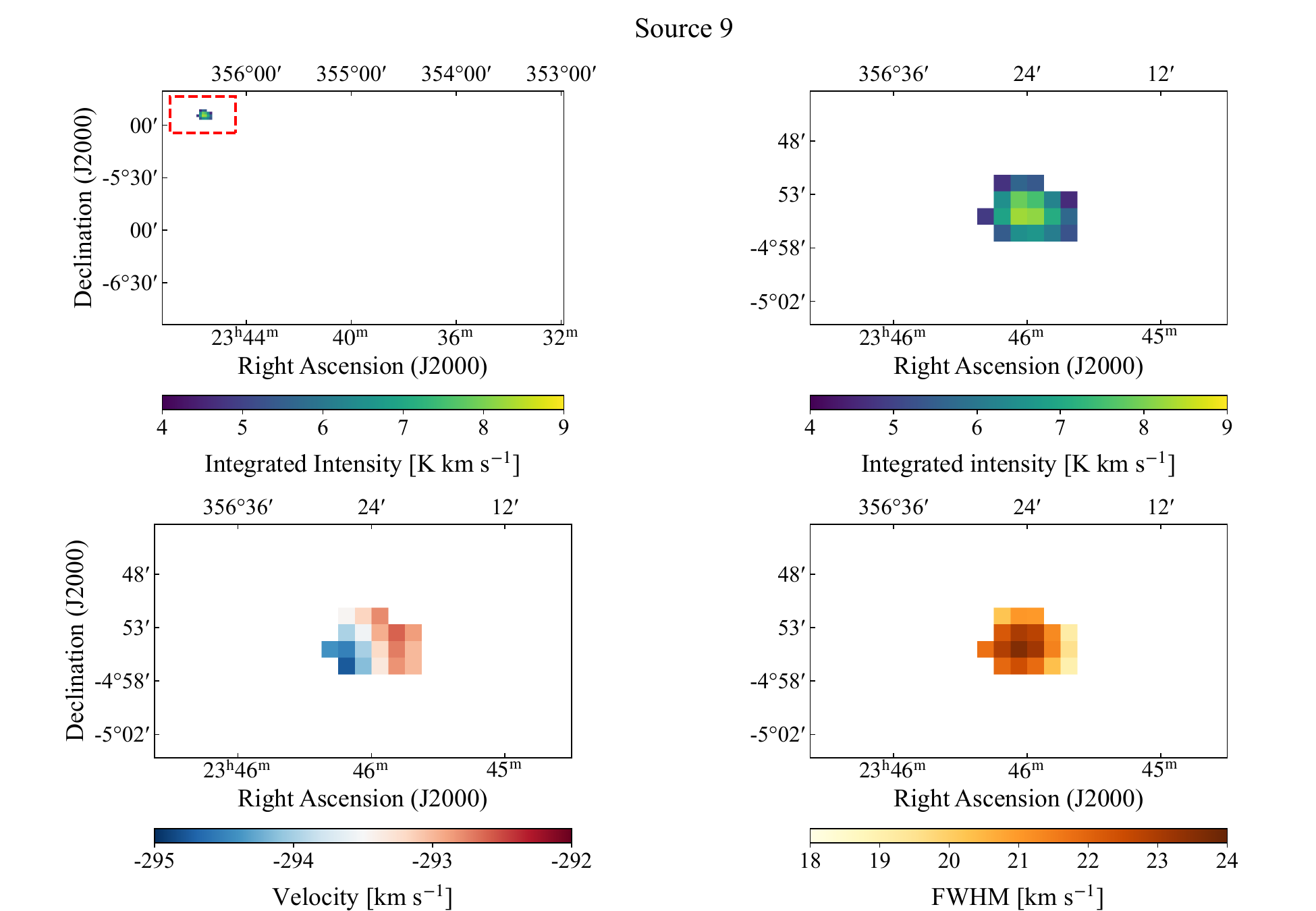}
\addtocounter{figure}{-1}
\caption{Continued.}
\end{figure*}



\end{CJK*}
\end{document}